\documentclass[iop,apj]{emulateapj}

\pdfoutput=1

\usepackage{times}
\usepackage[UKenglish]{isodate}    
\usepackage{multirow}
\usepackage{color}
\usepackage[fleqn]{amsmath}
\usepackage{amssymb}

\cleanlookdateon    

\defcitealias{harris1996}{H96}
\defcitealias{casetti2013}{CD13}

\newcommand {\afe} {\rm [\alpha/Fe]}
\newcommand {\feh} {\rm [Fe/H]}

\newcommand {\Rtidal} {R_{\rm tidal}}

\newcommand {\mua} {\mu_{\alpha*}}
\newcommand {\mud} {\mu_\delta}

\newcommand {\mi} {\mathbf{m}_{\rm i}}
\newcommand {\Ci} {\mathbf{C}_{\rm i}}
\newcommand {\alphai} {\alpha_{\rm i}}
\newcommand {\deltai} {\delta_{\rm i}}

\newcommand {\mGC} {\mathbf{m}_{\rm GC}}
\newcommand {\CGC} {\mathbf{C}_{\rm GC}}
\newcommand {\alphaGC} {\alpha_{\rm GC}}
\newcommand {\deltaGC} {\delta_{\rm GC}}

\newcommand {\Lppi} {\mathcal{L}_{\varpi\mu,\rm i}}
\newcommand {\Ladi} {\mathcal{L}_{\alpha\delta,\rm i}}
\newcommand {\Li} {\mathcal{L}_{\rm i}}

\newcommand {\vrad} {v_{\rm rad}}
\newcommand {\vtan} {v_{\rm tan}}

\definecolor{grey}{rgb}{0.5,0.5,0.5}

\usepackage{hyperref}
\usepackage[all]{hypcap}

\hypersetup{
    breaklinks,
    colorlinks,
    citecolor=blue,
    linkcolor=blue,
    urlcolor=blue,
    pdftitle={Proper motions in Galactic globular clusters with TGAS},
    pdfauthor={Laura L. Watkins, et al.}
}

\def\equationautorefname~#1\null{%
  equation~(#1)\null
}
\begin{document}

\title{Tycho-\textit{Gaia} Astrometric Solution parallaxes and proper motions for five Galactic globular clusters}

\author{Laura~L.~Watkins and Roeland~P.~van~der~Marel}
\affil{Space Telescope Science Institute, 3700 San Martin Drive, Baltimore MD 21218, USA}
\email{lwatkins@stsci.edu}

\slugcomment{ApJ draft, \today}
\shorttitle{TGAS parallaxes and proper motions for five Galactic globular clusters}
\shortauthors{Watkins \& van der Marel}


\begin{abstract}
    We present a pilot study of Galactic globular cluster (GC) proper motion (PM) determinations using \textit{Gaia} data. We search for GC stars in the Tycho-\textit{Gaia} Astrometric Solution (TGAS) catalogue from \textit{Gaia} Data Release 1 (DR1), and identify five members of NGC\,104 (47\,Tucanae), one member of NGC\,5272 (M\,3), five members of NGC\,6121 (M\,4), seven members of NGC\,6397, and two members of NGC\,6656 (M\,22). By taking a weighted average of member stars, fully accounting for the correlations between parameters, we estimate the parallax (and, hence, distance) and PM of the GCs. This provides a homogeneous PM study of multiple GCs based on an astrometric catalogue with small and well-controlled systematic errors and yields random PM errors similar to existing measurements. Detailed comparison to the available \textit{Hubble Space Telescope} (\textit{HST}) measurements generally shows excellent agreement, validating the astrometric quality of both TGAS and \textit{HST}. By contrast, comparison to ground-based measurements shows that some of those must have systematic errors exceeding the random errors. Our parallax estimates have uncertainties an order of magnitude larger than previous studies, but nevertheless imply distances consistent with previous estimates. By combining our PM measurements with literature positions, distances, and radial velocities, we measure Galactocentric space motions for the clusters and find that these also agree well with previous analyses. Our analysis provides a framework for determining more accurate distances and PMs of Galactic GCs using future \textit{Gaia} data releases. This will provide crucial constraints on the near end of the cosmic distance ladder and provide accurate GC orbital histories.
\end{abstract}



\section{Introduction}
\label{sect:introduction}

Accurate distances and space motions of Galactic globular clusters (GCs) are difficult but incredibly valuable to measure: improved constraints will have implications for the origin and evolution of the clusters themselves, for the structure of the Milky Way (MW), and for the fine-tuning of cosmological models.

The prevailing theory for the origin of the MW's GC population is that some GCs were formed outside of the MW and later accreted while the rest were formed in situ \citep[e.g.,][]{mackey2004}. The different origins and histories for the accreted and in-situ populations will manifest in their orbits \citep[e.g.,][]{zhu2016} and imprint on their internal dynamics \citep[e.g.,][]{webb2014, zocchi2016}, so determining which GCs belong to which population and analysing the differences between them will advance our understanding of cluster formation and evolution.

Moreover, accurate six-dimensional phase-space information for the GCs will allow us to constrain cluster orbits \citep[e.g.,][]{kuepper2015}, which will indicate which clusters have been most affected by tides and if any known GCs could be the missing progenitors of tidal streams whose origin is unknown. Cluster orbits will also be beneficial in constraining the inner shape of the Galactic halo \citep[e.g.,][]{pearson2015}.

Local GCs populate the short end of the cosmological distance ladder, so improved distances will provide important cosmology-independent verification for cosmological models \citep{verde2013}. Distances are also a key ingredient in GC age determination using the luminosity of the main-sequence (MS) turn off \citep[e.g.][]{chaboyer1995} and currently, the distance uncertainties in such analyses dominate over other sources of error. Accurate GC ages are of special interest as GCs are among the oldest objects in the universe for which ages are known and so help to constrain the age of the universe \citep[e.g.,][]{
krauss2003}.

Not only are distances and proper motions (PMs) useful measurements to have, but multiple measurements, using different data types and analysis methods, improve accuracy and ensure that the measurements are not biased. GC distances are typically estimated photometrically using ``standard candles'', such as RR Lyraes or the position of the horizontal branch (HB). However, it is also possible to estimate distances from kinematic data by comparing line-of-sight velocity and PMs measurements. In \citet{watkins2015b}, we estimated dynamical distances for 15 Galactic GCs using the \textit{Hubble Space Telescope (HST)} Proper Motion (HSTPROMO)\footnote{\url{http://www.stsci.edu/~marel/hstpromo.html}} GC catalogues \citep{bellini2014} and compared them to the photometric distance estimates compiled in \citet[][2010 edition, hereafter \citetalias{harris1996}]{harris1996}. Overall we showed that the two very different methods of distance estimation agreed very well, implying that both methods are robust and unbiased, however, a further comparison against parallax distances will provide a useful additional test.

Absolute PMs have previously been measured from the ground for 63 Galactic GCs \citep[][hereafter \citetalias{casetti2013}]{dinescu1997, dinescu1999a, dinescu1999b, dinescu2003, casetti2007, casetti2010, casetti2013}\footnote{A compilation is available at \url{http://www.astro.yale.edu/dana/gc.html}.} as part of the Southern Proper Motion (SPM) Program. These measurements are heterogeneous as the method used to correct the PMs to an inertial reference frame varied from cluster to cluster. Also, in some cases, the measurements represent a combination of multiple determinations (by including other PM determinations from other ground- and space-based sources) and not a single determination. Nevertheless, independent PM measurements for these GCs will be incredibly helpful to verify their accuracy and to assess the consistency of the different methods.

\textit{Gaia} will provide five-parameter astrometric solutions -- positions, parallaxes, and PMs -- for objects brighter than $\sim 20$~mag, so it will be tremendously useful for analysing both internal and global PMs for Galactic GCs and their distances\footnote{Note that \citet{astraatmadja2016} caution that estimation of distances from \textit{Gaia} parallaxes is not as simple as inverting the parallaxes, since the parallax uncertainties are non-negligible and must be done using rigorous statistical analysis.}. PM and parallax are both degenerate, and a sufficiently long baseline is needed to disentangle the two; at the time of the first \textit{Gaia} data release \citep[DR1, 2016 September,][]{gaia2016}, parallaxes and PMs were not measurable with \textit{Gaia} data alone. The first \textit{Gaia}-only PM measurements are projected for release in late 2017 and the final (and most accurate) PMs will be released only in 2022. However, by crossmatching the \textit{Gaia} catalogue with the \textit{Hipparcos} Tycho2 catalogue \citep{hog2000}, it is possible to measure PMs (and parallaxes) for stars in both datasets: the Tycho-\textit{Gaia} Astrometric Solution \citep[TGAS,][]{michalik2015, lindegren2016}. Combining Tycho2 and \textit{Gaia} data together offers an extended baseline for PM studies that has made PM analyses achievable for a few million stars with the \textit{Gaia} DR1. Although not ideally suited to dynamical studies of objects within the Local Group, TGAS has already been used to measure the rotation fields of the Magellanic Clouds \citep{vandermarel2016} and to measure the space motion of GC NGC\,2419 \citep{massari2017}. Here, we search for Galactic GC stars in the TGAS catalogue.

This is a pilot study based on the limited information currently available. We hope that our methods can serve as a template for studies with future Gaia data releases, which will have more stars and will yield higher accuracy PM results.

We outline our cluster-member determination in \autoref{sect:membership}, present our parallax and PM results in \autoref{sect:ppms}, discuss the implied space motions and orbits of the clusters in \autoref{sect:spacemotion}, and conclude in \autoref{sect:conclusions}.


\section{Cluster membership determination}
\label{sect:membership}

To determine likely GC members, we search for stars that are close to a GC centre on the plane of sky and then check for consistency with previous PM and parallax estimates and typical GC isochrones. We also wish to consider predictions for the PMs, parallaxes, and photometry expected for MW stars along the line of sight of the GC to rule out contaminants.

\subsection{Proximity}
\label{ssect:proximity}

To estimate how many Galactic GC stars may exist in the TGAS catalogue, we count the number of stars found within some limiting radius from the centre of each GC on the plane of the sky. We begin with the \citetalias{harris1996} Galactic GC catalogue from which we extract cluster centre coordinates, concentrations, and core radii. Although the catalogue contains data for 157 Galactic GCs, we were only able to proceed with 156 GCs, since Pyxis has no concentration or core radius estimate listed.\footnote{This does not affect our analysis, as Pyxis is too distant to have stars with sufficient apparent brightness in the Tycho2 catalogue.}

We use the concentrations $c$ and core radii $R_{\rm core}$ to estimate the tidal radii $\Rtidal$ of the GCs via
\begin{equation}
    \Rtidal = 10^c R_{\rm core}.
\end{equation}
This is not generally considered to be a robust method of tidal-radius determination and is likely to underestimate the true extent of a GC. These tidal radii estimates assume an underlying King profile, however, GCs are generally better fit by Wilson models \citep[e.g.,][]{mclaughlin2005}, which are still finite in extent but are more extended than King models. Furthermore, a GC that is tidally disrupting may have member stars outside of its formal tidal radius. However, this method is sufficient for our purposes, which is simply to estimate the approximate extent of the clusters on the plane of the sky. To mitigate the possibility of underestimating the true extent of the cluster, we adopt $2 \Rtidal$ as the limiting radius within which to search for GC members.

Then, for each GC in turn, we extracted all stars within $2 \Rtidal$ of the cluster centre from the TGAS catalogue. We did this for all 156 GCs with no cuts on heliocentric distance, magnitude, or central velocity dispersion (all of which could affect the likelihood that a star close to the centre in projection is a true cluster member). In total, we identified 4268 stars within $2 \Rtidal$ of the centre of a GC across 142 clusters.

\subsection{Distance and Extinction}
\label{ssect:dmext}

Some of these clusters are relatively far away and others are in regions of high extinction, so it is likely that true GC members will be rendered too faint for detection with TGAS. So we wish to determine the expected apparent magnitude of the brightest star in any given GC and then the number of observed stars fainter than this limit.

To do this, we select the $V_{\rm T}$ magnitude from the Tycho2 catalogue or the $H_p$ magnitude from the Hipparcos catalogue for each star, depending on the original catalogue of origin. Although there is not a perfect correspondence, both Hipparcos $H_p$ and Tycho2 $V_{\rm T}$ can be approximated by the \textit{HST} F555W filter, for which we have isochrones and extinction coefficients. We then take a typical GC isochrone from the Dartmouth Stellar Evolution Database \citep{dotter2008} with metallicity $\feh = -1.5$~dex, alpha-element abundance $\afe = 0.2$~dex, and age $A = 11$~Gyr, and estimate the brightest magnitude that the isochrone reaches.

For each cluster, we adjust the tip magnitude of the isochrone for GC distance and extinction along the line of sight using distance moduli calculated using distances from \citetalias{harris1996}, extinctions also from \citetalias{harris1996}, and extinction coefficients from \citet{sirianni2005}. To allow for magnitude uncertainties, we set the limiting magnitude at 0.5~mag brighter than the distance- and extinction-corrected tip of the isochrone and count the number of stars fainter than this limiting magnitude. We perform a more careful photometry check on our GC member candidates in \autoref{ssect:photometry}, this rather crude preliminary cut is intended to remove distant and highly extincted clusters from further consideration. After these cuts, we are left with 967 stars across 30 clusters; the most distant of these clusters lies at 10.3~kpc.

\subsection{Parallax and Proper Motion}
\label{ssect:probs}

To determine which of these stars are likely to be cluster members, we wish to identify which stars have parallax and PM measurements that are consistent with previous measurements for their nearby cluster.

TGAS provides parallaxes $\varpi$ and PMs $(\mua, \mud)$ for each star, along with uncertainties for each and correlation coefficients among the parameters, from which we can construct a full 3-dimensional covariance matrix.

For the GCs, we adopt distances from \citetalias{harris1996}, which we then invert to determine parallax estimates. The \citetalias{harris1996} distances have no formal error bars, so we adopt errors of 10\% on our parallax estimates. For PM measurements, we use values from the compilation described in \citetalias{casetti2013} and earlier papers, which themselves come from a variety of sources and are sometimes averages of multiple measurements. Only 26 of the 30 clusters with GC member candidates have PM measurements in this compilation and so only these 26 GCs can be evaluated here.

For each cluster, we calculate the probability for the nearby stars of being a cluster member. For a star $i$ with parallax and PM measurements $\mi$ and covariance $\Ci$, we ask what the likelihood $\Lppi$ is that this star is a member of a GC with measurements $\mGC$ and covariance $\CGC$,
\begin{align}
    \Lppi & = p \left( \mi \, | \, \Ci, \mGC, \CGC \right) \nonumber \\
    & = \frac{ \exp \left[ - \frac{1}{2} \left( \mi - \mGC \right)^{\mathrm{T}} \left( \Ci + \CGC \right)^{-1} \left( \mi - \mGC \right) \right] } { \sqrt{ \left( 2 \mathrm{\pi} \right)^3 \left| \left( \Ci + \CGC \right) \right| } },
\end{align}
which is a standard 3-dimensional Gaussian. To construct the GC covariance matrix $\CGC$, we assume that the errors are uncorrelated, so the diagonal terms are the squared uncertainties on the parallax and PM measurements and the off-diagonal elements are zero. We further add the GC velocity dispersion \citepalias{harris1996} in quadrature to the PM terms to account for the expected spread in velocities.

This likelihood calculation assumes that all stars selected near the cluster are equally likely to be members, but this is not the case. Stars close to the cluster centre of the cluster are more likely to be members than stars near to the $2 \Rtidal$ boundary, so we also calculate the likelihood of a star $i$ with coordinates $(\alphai, \deltai)$ being a member of the GC with centre $(\alphaGC, \deltaGC)$ as
\begin{align}
    \Ladi & = p \left( \alphai, \deltai \, | \, \alphaGC, \deltaGC, \sigma \right) \nonumber \\
    & = \exp \left[ - \frac{1}{2 \sigma^2} \left( \left( \alphai - \alphaGC \right)^2 + \left( \deltai - \deltaGC \right)^2 \right) \right].
\end{align}
where we use $\sigma = \frac{1}{2} \Rtidal$ to account for the approximate extent of the cluster. The uncertainties of the cluster centre coordinates and of the positions of the stars are negligible compared to the extent of the cluster, so we neglect the measurement errors in this calculation.

GC projected number density profiles are, of course, highly non-Gaussian, however, the profiles in the outer regions of GCs are largely uncertain due to effect of tides and the scarcity of data there. This Gaussian weighting suffices to down-weight stars found further away from the cluster centre. Also note that we have used unnormalised Gaussians here, as we are mostly concerned with how many tidal radii the star is away from the cluster centre, rather than the tidal radius itself. That is, a star 1$\Rtidal$ from the cluster centre should have the same $\Ladi$ in all clusters, regardless of the value of $\Rtidal$.

Finally, we can calculate the total likelihood that the star is a member of the cluster $\Li = \Lppi \times \Ladi$. We keep all stars with $\Li > -11$ as possible cluster members.\footnote{We experimented with different likelihood cuts and found that the analysis is fairly insensitive to the precise likelihood cut value but that a cut at $\Li > -11$ works well.} This selects 64 stars across 15 clusters. All of the identified candidates are Tycho2 stars, no Hipparcos stars remain in our sample.

\subsection{Radial Velocities}
\label{ssect:radvel}

As a further test for membership, we check whether the candidate cluster members have radial velocities (RVs) consistent with cluster membership. We cross-match our stars with the RAdial Velocity Experiment \citep[RAVE,][]{kunder2016} catalogue to extract RVs and find 15 stars in common between our high-likelihood sample and RAVE. Using RV and velocity dispersion estimates from \citetalias{harris1996}, we sigma-clip the sample at 3$\sigma$ and reject five stars.

This leaves 59 candidate cluster members across 15 clusters: 11 stars in NGC\,6838 (M\,71); 8 stars in each of NGC\,104 (47\,Tuc) and NGC\,6752; 7 stars in NGC\,6397; 5 stars in each of NGC\,5139 ($\omega$ Centauri), NGC\,6121 (M\,4) and NGC\,6656 (M\,22); 2 stars in each of NGC\,6218 (M\,12) and NGC\,6254; and 1 star only in each of NGC\,4833, NGC\,5272 (M\,3), NGC\,6362, NGC\,6723 (M\,19), NGC\,6779 (M\,56), and NGC\,6809.

\subsection{Photometry}
\label{ssect:photometry}

\begin{figure}
    \centering
    \includegraphics[width=0.75\linewidth]{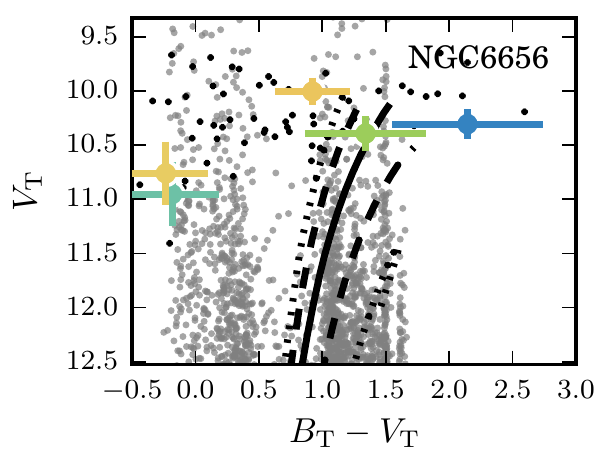}
    \caption{Representative distance-corrected isochrones for NGC\,6656. The solid line shows an 11~Gyr, $\afe$=0.2~dex isochrone with $\feh$ metallicity from \citetalias{harris1996} rounded to the nearest 0.5~dex, the dashed lines show the isochrones at $\pm$0.5~dex, and the dotted lines show the isochrones at $\pm$1~dex. The extinction-corrected TGAS cluster-member candidates are shown as coloured points where the colour indicates their likelihood from high (orange) to low (blue). The black points are TGAS stars ruled out as cluster members. For comparison, we show the expected MW field-star population from Besan\c{c}on model predictions as grey points. For this cluster, we reject the two stars with $B_{\rm T} - V_{\rm T} < 0$, but consider the others to be candidate cluster members.}
    \label{fig:cmds}
\end{figure}

Many of the identified clusters are classified as disk or bulge clusters, so it is likely that some stars identified as possible members are disk or bulge stars along the line of sight to the cluster. Furthermore, it is likely that high reddening would have made cluster stars too faint for detection with Tycho2. So the next step is to assess whether the stars are photometrically consistent with their nearest cluster. To do this, we wish to investigate whether the identified stars are consistent with an old isochrone at the known distance, reddening, and metallicity.

Once again, we cross-match with the Tycho2 catalogue to extract $B_{\rm T}$ and $V_{\rm T}$ magnitudes, which are very close to the Johnson $B$ and $V$ magnitudes. We then extract representative isochrones for each cluster from the Dartmouth Stellar Evolution Database \citep{dotter2008}. To do this, we assume an $\alpha$-element abundance $\afe$ of 0.2~dex and an age of 11~Gyr, and use $\feh$ metallicities from \citetalias{harris1996} rounded to the nearest 0.5~dex. We adjust the isochrone magnitudes for distance \citepalias[using distance moduli calculated from distances in][]{harris1996}, and correct the observed magnitudes for extinction using reddening values from \citetalias{harris1996} and extinction coefficients from \citet{sirianni2005}, using the F435W filter to approximate $B_{\rm T}$ and the F555W filter to approximate $V_{\rm T}$.\footnote{These approximations are sufficient for present analysis, which does not rely on accurate photometry.}

As an example, in \autoref{fig:cmds}, we show distance-corrected isochrones for NGC\,6656. Results for the other 14 clusters are given in \autoref{sect:extrafigs}. The solid line shows the isochrone with cluster $\feh$, the dashed lines show the isochrones at $\pm$0.5~dex, and the dotted lines show the isochrones at $\pm$1~dex. The TGAS stars identified as possible members are shown as coloured points, with their colours representing their likelihood of membership based on their position, parallax, and PM, from orange (high) to blue (low). TGAS stars already ruled out as members (and so likely MW foreground stars) are shown in black.

From a visual inspection of the resulting isochrones for each cluster, we find that the candidate cluster members in NGC\,4833, NGC\,6362, NGC\,6723, and NGC\,6779 are all significantly offset from the isochrones and so we remove these from further consideration. We also reject 1 star in NGC\,6254, 2 stars in NGC\,6656, 1 star in NGC\,6752, and 3 stars in NGC\,6838 that are all significantly bluer than the expected cluster isochrones. The other stars remain promising candidates, and we still have 48 possible members across 11 clusters.

\subsection{Besan\c{c}on Simulations}
\label{ssect:besancon}

As a final test, we wish to compare the velocities of the TGAS stars against predictions from the Besan\c{c}on simulations \citep{robin2003}; these simulations serve as a prediction for what the population of field stars should look like -- both photometrically and kinematically -- along the line of sight to any given cluster.

\begin{figure}
    \centering
    \includegraphics[width=\linewidth]{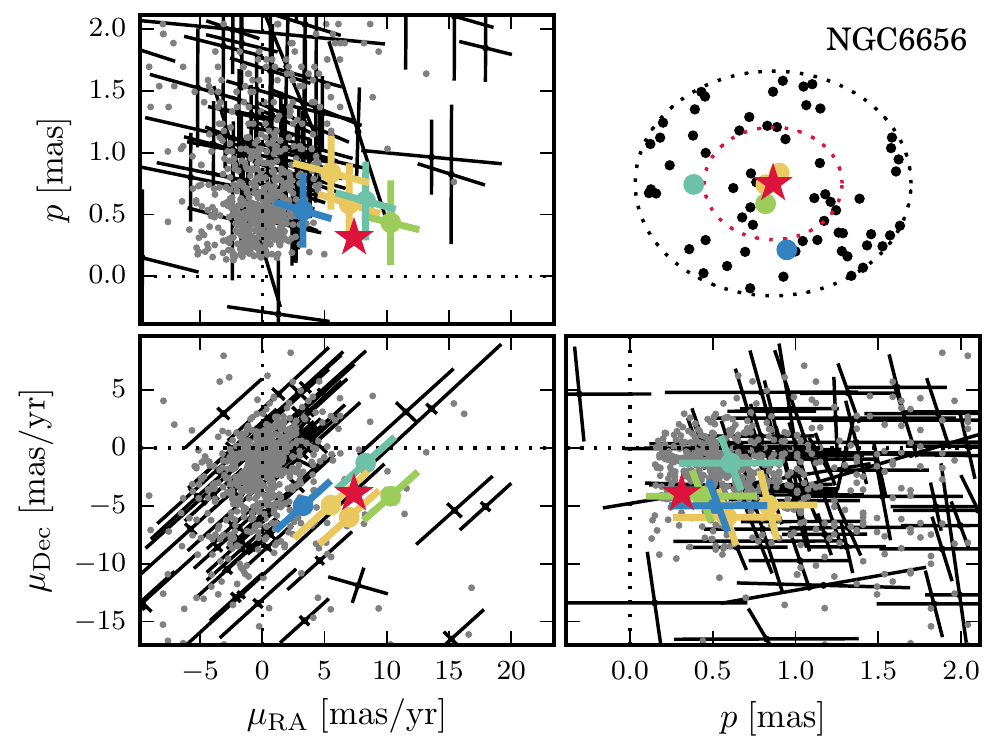}
    \caption{Sky positions, parallaxes, and proper motions for the TGAS stars in NGC\,6656,. The top-right panel shows the sky positions of the TGAS stars identified within $2 \Rtidal$ relative to the cluster centre (red star). The red dotted line marks $\Rtidal$ and the black dotted line marks $2 \Rtidal$. The remaining panels show projections of the five-parameter astrometric solution for parallax, RA PM and Dec PM. The literature values we used to determine membership probabilities (parallax from the inverted distance from \citetalias{harris1996} and PMs from \citetalias{casetti2013}) are shown as red stars. The TGAS cluster-member candidates are shown as coloured points, coloured by their probability of membership from orange (high) to blue (low). The black points are TGAS stars rejected as cluster members, and thus, thought to be MW field stars. For comparison, Besan\c{c}on model predictions are shown as grey points. Some of the Besan\c{c}on stars fall outside of the plot limits.}
    \label{fig:probs}
\end{figure}

We selected stars from the Besan\c{c}on simulations using the webtool provided, and we found that all of the default inputs were suitable for our purposes. We used the small-field option to select stars and used Galactic longitudes and latitudes from \citetalias{harris1996} to identify the cluster centres.

As an example, in \autoref{fig:probs}, we show 2D projections of the parallax, RA PM and Dec PM space for NGC\,6656. Results for the other 10 clusters are given in \autoref{sect:extrafigs}. The red stars indicate the \citetalias{harris1996} distance inverted to calculate a parallax and the PMs from \citetalias{casetti2013}. The coloured points are the candidate cluster members from TGAS, coloured according to their probability of membership from orange (high) to blue (low). The black points are TGAS stars selected inside $2 \Rtidal$ rejected as cluster-member candidates (and so likely MW field stars). The grey points show the predictions for the MW field population from Besan\c{c}on simulations. Note that the error bars are not aligned with the axes, as we have incorporated the correlation terms in the five-parameter astrometric solution provided in the TGAS catalogue. The error bars show the major and minor axes of the 2D projection of the full covariance matrix. In the top-right corners, we show sky position maps for the TGAS candidate cluster members (coloured points) and the TGAS field stars (black points) relative to the centre of the cluster (red star). The red dotted lines marks the tidal radii $\Rtidal$, and the black dotted lines mark $2 \Rtidal$.

We find that both the photometric and kinematic properties for 6 of the remaining 11 clusters are consistent with the Besan\c{c}on predictions for the field-star population, and the other TGAS stars in the respective fields, which are also assumed to be field stars. In all cases, we cannot be confident that we have truly selected cluster members and not field stars, and so we remove these clusters from further consideration. The removed clusters are: NGC\,5139, NGC\,6218, NGC\,6254, NGC\,6752, NGC\,6809, and NGC\,6838.

Thus, we are left with five promising clusters, which we now discuss in turn.

\textbf{NGC\,104} has eight candidate stars, but it also poses the challenge that the Besan\c{c}on model predictions overlap the part of parallax and PM space where NGC\,104 stars are expected. Therefore, for this cluster we decided to restrict our final analysis to only the stars within $0.5 \Rtidal$ from the centre, which is where the cluster-to-foreground star ratio should be largest. There are five stars that meet this criterion, and all of them clump in PM space around the expected NGC\,104 PM value. While we cannot strictly rule out that any of these five stars are foreground, we can rule out that a significant fraction of them are foreground, since the Besan\c{c}on models predict a much larger PM dispersion for such stars. One of the five candidate member stars also has an RV estimate within 3$\sigma$ of the cluster value (in fact, at 1.2$\sigma$). This does lend further credence to their membership (although we find that 65\% of the Besan\c{c}on stars also have RVs predicted to be within 3$\sigma$ of the cluster value).

\textbf{NGC\,5272} has only a single candidate, however, it is very close to the cluster centre and the agreement in PM and parallax is excellent. Furthermore, both cluster and star sit far away from the Besan\c{c}on predictions. We believe that this star is a member of NGC\,5272, though we acknowledge that it is hard to be certain for a single star.

\textbf{NGC\,6121} has five candidates; all lie well within the tidal radius and all have PMs and parallaxes consistent both with the cluster and each other and are inconsistent with the Besan\c{c}on predictions. We note that one star is slightly offset from the others, however, its proximity to the centre and its position relative to a cluster isochrone both mark it as a promising candidate. Unfortunately, this offset star is not augmented by RV data, however, two other candidates do have RV estimates: one lies at $\sim 0.2\sigma$ from the cluster value and the other at $\sim 1.1\sigma$ from the cluster value, thus indicating that they are indeed bona fide cluster members. We believe all five stars are cluster members.

\textbf{NGC\,6397} has seven candidates, all of which are clearly offset from the bulk of the Besan\c{c}on predictions in PM. One star (light blue) also has an RV measurement within 1$\sigma$ of the cluster measurement, however, $\sim$22\% of the Besan\c{c}on stars have RVs within 3$\sigma$ of the cluster value, so RV is not a strong indicator of cluster membership. Nevertheless, we believe that all seven stars are members.

\textbf{NGC\,6656} has three candidates, all of which have parallaxes and $\mud$ measurements consistent with the Besan\c{c}on population; two (yellow and green) are clearly offset in $\mua$ from the bulk of the background predictions but the third (blue) does sit at the edge of the predicted background distribution; as this star also sits outside of the tidal radius and is somewhat redward of the expected cluster isochrones, we remove it from our analysis. Of the remaining two stars, the star coloured yellow has an RV measurement that is 2$\sigma$ offset from the cluster value. Fewer than 3\% of the Besan\c{c}on stars have RVs within 3$\sigma$ of the cluster velocity, so RV is indeed a strong indicator of cluster membership for this cluster. We believe that the remaining two stars are members.

\begin{table*}
    \caption{TGAS star properties and membership selection statistics.}
    \label{table:stars}
    \centering
    \renewcommand{\arraystretch}{0.01}
    
    \begin{tabular}{cccccccccccccc}
        \hline
        \hline
        GID & $\alpha$ & $\delta$ & $\varpi$ & $\mua$ & $\mud$ & $B_{\rm T}$ & $V_{\rm T}$ & $v_{\rm r}$ & Cluster & $\log \Ladi$ & $\log \Lppi$ & $\log \Li$ &  \\
        & \multicolumn{2}{c}{(deg)} & (mas) & \multicolumn{2}{c}{(mas/yr)} & \multicolumn{2}{c}{(mag)} & (km/s) &  &  &  &  &  \\
        (1) & (2) & (3) & (4) & (5) & (6) & (7) & (8) & (9) & (10) & (11) & (12) & (13) &  \\
        \hline \\[-1em]
        4689620330317403136 & 5.91 & -72.28 & 0.274 & 4.22 & -2.73 & 13.82 & 12.01 &   \dots & NGC\,104 & -0.204 & -2.478 &  -2.683 & $\star$ \\[0.4em]
        4690024022888359424 & 7.30 & -71.82 & 0.109 & 3.96 & -1.60 & 13.64 & 11.72 &  -19.28 & NGC\,104 & -6.887 & -3.748 & -10.634 &         \\[0.4em]
        4689832845301844352 & 6.09 & -71.89 & 1.058 & 5.06 & -4.97 & 14.48 & 11.74 &   \dots & NGC\,104 & -0.165 & -4.036 &  -4.201 & $\star$ \\[0.4em]
        4689644416501132800 & 6.22 & -71.94 & 0.375 & 7.10 & -4.50 & 13.02 & 12.26 &   \dots & NGC\,104 & -0.235 & -3.745 &  -3.980 & $\star$ \\[0.4em]
        4689638437899435136 & 5.57 & -72.10 & 0.580 & 2.09 & -1.14 & 13.28 & 12.20 &   \dots & NGC\,104 & -0.814 & -3.447 &  -4.260 &         \\[0.4em]
        4689645000616682240 & 6.01 & -71.93 & 0.757 & 4.01 & -3.32 & 13.28 & 11.65 &  -30.57 & NGC\,104 & -0.093 & -3.006 &  -3.099 & $\star$ \\[0.4em]
        4689623594492482176 & 6.27 & -72.16 & 0.180 & 7.14 & -4.43 & 12.71 & 11.55 &   \dots & NGC\,104 & -0.259 & -4.033 &  -4.292 & $\star$ \\[0.4em]
        4689595831823970304 & 5.41 & -72.41 & 0.675 & 4.24 & -3.25 & 13.54 & 12.00 &   \dots & NGC\,104 & -1.944 & -3.268 &  -5.212 &         \\[0.4em]
        \hline
        1454784965950282240 & 205.46 & 28.39 & 0.225 & -0.69 & -2.85 & 14.74 & 12.46 & -148.86 & NGC\,5272 & -0.062 & -0.939 &  -1.001 & $\star$ \\[0.4em]
        \hline
        6045476290889048704 & 245.82 & -26.54 & 0.586 & -12.78 & -18.38 & 14.13 & 11.65 &   69.90 & NGC\,6121 & -0.017 & -2.094 &  -2.111 & $\star$ \\[0.4em]
        6045462100309586816 & 245.97 & -26.58 & 0.437 & -12.56 & -18.75 & 13.10 & 11.68 &   \dots & NGC\,6121 & -0.023 & -2.018 &  -2.042 & $\star$ \\[0.4em]
        6045462890583403136 & 245.84 & -26.62 & 0.425 & -12.87 & -18.51 & 11.99 & 10.15 &   \dots & NGC\,6121 & -0.033 & -1.977 &  -2.011 & $\star$ \\[0.4em]
        6045490618893190656 & 246.01 & -26.40 & 0.521 & -13.41 & -18.88 & 12.86 & 10.94 &   66.26 & NGC\,6121 & -0.078 & -2.061 &  -2.138 & $\star$ \\[0.4em]
        6045452204710238080 & 245.86 & -26.79 & 0.810 & -10.95 & -17.14 & 12.93 & 10.91 &   \dots & NGC\,6121 & -0.189 & -5.091 &  -5.280 & $\star$ \\[0.4em]
        \hline
        5921742474972627456 & 265.18 & -53.83 & 0.317 & 2.28 & -18.92 & 11.92 & 10.54 &   \dots & NGC\,6397 & -0.731 & -5.469 &  -6.200 & $\star$ \\[0.4em]
        5921747938171016320 & 265.08 & -53.70 & -0.150 & 3.45 & -17.91 & 11.79 & 10.18 &   \dots & NGC\,6397 & -0.260 & -4.060 &  -4.319 & $\star$ \\[0.4em]
        5921745567349079424 & 265.11 & -53.80 & 0.441 & 2.04 & -19.45 & 12.64 & 12.06 &   \dots & NGC\,6397 & -0.587 & -5.876 &  -6.463 & $\star$ \\[0.4em]
        5921729074674635776 & 265.47 & -53.56 & 0.159 & 3.79 & -17.58 & 12.06 & 10.68 &   \dots & NGC\,6397 & -2.945 & -7.572 & -10.517 & $\star$ \\[0.4em]
        5921744708355615616 & 265.18 & -53.75 & 0.098 & 4.61 & -17.91 & 11.96 & 10.55 &   \dots & NGC\,6397 & -0.148 & -6.352 &  -6.500 & $\star$ \\[0.4em]
        5921751030547464960 & 265.24 & -53.66 & -0.041 & 3.14 & -16.24 & 12.41 & 11.20 &   23.21 & NGC\,6397 & -0.120 & -8.766 &  -8.887 & $\star$ \\[0.4em]
        5921744570916663040 & 265.16 & -53.76 & 0.608 & 1.95 & -17.19 & 11.77 & 10.54 &   \dots & NGC\,6397 & -0.202 & -4.154 &  -4.357 & $\star$ \\[0.4em]
        \hline
        4077494753703271424 & 279.04 & -23.92 & 0.585 & 6.98 & -5.99 & 12.31 & 11.09 & -130.00 & NGC\,6656 & -0.023 & -4.215 &  -4.238 & $\star$ \\[0.4em]
        4076740385647365376 & 279.04 & -24.10 & 0.433 & 10.31 & -4.17 & 13.12 & 11.47 &   \dots & NGC\,6656 & -0.285 & -5.788 &  -6.073 & $\star$ \\[0.4em]
        4077552649862421248 & 278.49 & -23.91 & 0.609 & 8.30 & -1.32 & 12.16 & 12.03 &   \dots & NGC\,6656 & -2.650 & -4.648 &  -7.298 &         \\[0.4em]
        4077590892251201152 & 279.15 & -23.81 & 0.837 & 5.52 & -4.94 & 11.91 & 11.84 &   \dots & NGC\,6656 & -0.085 & -4.411 &  -4.497 &         \\[0.4em]
        4076708431092287104 & 279.21 & -24.53 & 0.533 & 3.28 & -5.00 & 13.84 & 11.39 &   \dots & NGC\,6656 & -2.874 & -6.778 &  -9.652 &         \\[0.4em]
        \hline
    \end{tabular}
    
    \qquad
    
    \textbf{Notes.} Columns: (1) \textit{Gaia} ID; (2) Right Ascension; (3) Declination; (4) parallax; (5) PM in Right Ascension; (6) PM in Declination; (7) Tycho2 B magnitude; (8) Tycho2 V magnitude; (9) radial velocity from RAVE; (10) nearest cluster centre; (11) logarithm of position likelihood; (12) logarithm of parallax and PM likelihood; (13) logarithm of total likelihood. Stars marked $\star$ are those we identify as cluster members and use in \autoref{sect:ppms} to derive the cluster properties in \autoref{table:results}. Uncertainties and correlations are not listed, but are available from the TGAS, Tycho2 and RAVE catalogues.
    
\end{table*}

In summary, we are confident that we have identified 20 GC member stars: 5 stars in NGC\,104, 1 star in NGC\,5272, 5 stars in NGC\,6121, 7 stars in NGC\,6397, and 2 stars in NGC\,6656. In \autoref{table:stars}, we provide the TGAS and Tycho2 properties for all stars in these clusters that passed the parallax-PM likelihood cut, along with their known RVs, and the likelihoods calculated in \autoref{ssect:probs}. Stars marked with a $\star$ are those in the final sample that we believe to be bona fide members.

Perhaps unsurprisingly, NGC\,6121, NGC\,6397, and NGC\,6656 are three of the four closest clusters.\footnote{The fourth cluster is NGC\,6544, which was excluded from our analysis as it has no stars inside $2 \Rtidal$ that passed the distance and extinction test in \autoref{ssect:dmext}.} Thanks in part to their proximity, these clusters are among the best studied and so precise measurements of both their distances (parallaxes) and their absolute PMs are extremely valuable to aid our understanding of their nature and origins.


\section{Proper motions and parallaxes}
\label{sect:ppms}

With just a single star in NGC\,5272, there is little more we can do here; however, for the other clusters we can combine the TGAS estimates and uncertainties to calculate weighted average parallaxes and PMs for the clusters. To do this, we perform a Monte-Carlo sampling using the member stars identified in TGAS to estimate the cluster PM and parallax along with their covariance.

We also evaluate the probability of any identified member star being a foreground interloper by calculating the $\chi^2$ in parallax for the member stars relative to the literature value, and then the corresponding $p$-value (probability of a $\chi^2$ this high or higher occurring by chance). For all clusters, we find $p \gg 0.05$, which indicates that there is no statistical basis to assume that any of the stars in our final samples must be an interloper (although this can never be fully excluded).

All of our results are presented in \autoref{table:results} and are compared to results from previous studies in the following subsections. Comparing our parallax estimates to previous measurements is subtly non-trivial, since usually previous studies have estimated distances or distance moduli, so two types of measurements (along with their uncertainties) must be converted into the third type for comparison. The choice of which measurement type to use for comparison is somewhat arbitrary, and we choose to compare distances here, as cluster distances are commonly the quantity that we wish to know.

\begin{table*}
    \caption{TGAS parallax and proper motion results.}
    \label{table:results}
    \centering
    
    \begin{tabular}{ccccccccccc}
        \hline
        \hline
        Cluster & $\varpi$ & $\mua$ & $\mud$ & $\sigma \varpi$ & $\sigma \mua$ & $\sigma \mud$ & $\rho(\varpi,\mua)$ & $\rho(\varpi,\mud)$ & $\rho(\mua,\mud)$ & $p_{\rm parallax}$ \\
        & (mas) & (mas/yr) & (mas/yr) & (mas) & (mas/yr) & (mas/yr) &  &  &  &  \\
        (1) & (2) & (3) & (4) & (5) & (6) & (7) & (8) & (9) & (10) & (11) \\
        \hline \\[-1em]
        NGC\,104 & 0.531 &   5.50 &  -3.99 & 0.210 & 0.70 & 0.55 & -0.440 & -0.503 & -0.332 & 0.428 \\[0.4em]
        NGC\,5272 & 0.225 &  -0.69 &  -2.85 & 0.289 & 0.51 & 0.37 & -0.197 & -0.593 &  0.319 & 0.660 \\[0.4em]
        NGC\,6121 & 0.556 & -12.51 & -18.33 & 0.149 & 0.50 & 0.29 &  0.589 &  0.519 &  0.488 & 0.930 \\[0.4em]
        NGC\,6397 & 0.205 &   3.03 & -17.88 & 0.223 & 1.09 & 1.36 & -0.066 &  0.036 &  0.755 & 0.895 \\[0.4em]
        NGC\,6656 & 0.509 &   8.64 &  -5.09 & 0.222 & 1.49 & 1.45 & -0.194 & -0.160 &  0.935 & 0.630 \\[0.4em]
        \hline
    \end{tabular}
    
    \qquad
    
    \textbf{Notes.} Columns: (1) cluster name; (2) parallax estimate; (3) RA PM estimate; (4) Dec PM estimate; (5) uncertainty on parallax estimate; (6) uncertainty on RA PM estimate; (7) uncertainty on Dec PM estimate; (8) correlation between parallax and RA PM estimates; (9) correlation between parallax and Dec PM estimates; (10) correlation between RA PM and Dec PM estimates; (11) $p$-value for parallax.
    
\end{table*}

\subsection{NGC\,104 (47\,Tucanae)}
\label{ssect:ppm104}

\begin{figure}
    \centering
    \includegraphics[width=\linewidth]{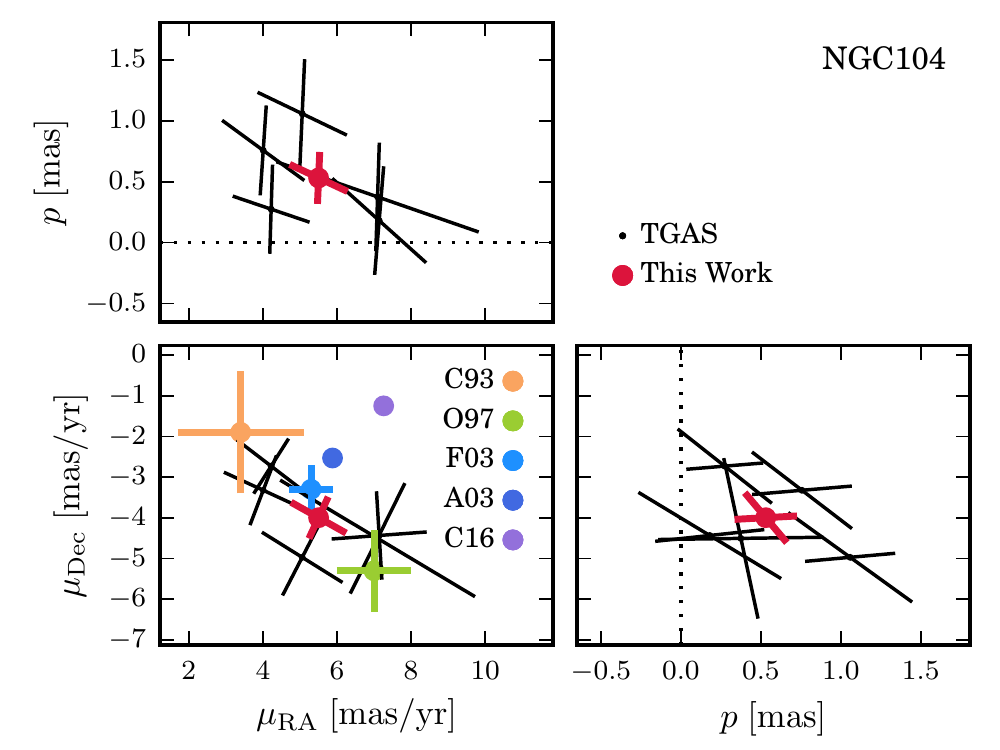}
    \includegraphics[width=\linewidth]{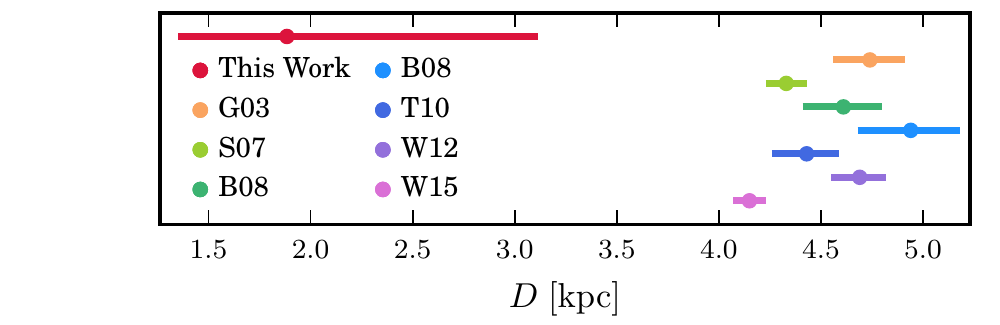}
    \caption{Upper and middle panels: parallax and PM results for NGC\,104. The black points show the five TGAS stars that we identified as cluster members. The red points show the mean values that we have calculated here. In the PM plot (middle left) we compare our estimate to values from \citet[orange, ground-based]{cudworth1993}, \citet[green, \textit{Hipparcos}]{odenkirchen1997}, \citet[cyan, millisecond pulsars]{freire2003}, \citet[blue, \textit{HST}]{anderson2003}, and \citet[purple, VISTA]{cioni2016}. Our PM estimate is consistent with both the estimate from MSP timing and the estimate from \textit{HST} astrometry. Lower panel: distance comparison for NGC\,104. We invert our parallax measurement to obtain a distance estimate (red). We also show previous distance estimates from \citet[orange, MS fitting]{gratton2003}, \citet[light green, HB fitting]{salaris2007}, \citet[dark green, TRGB; cyan RR Lyraes]{bono2008}, \citet[blue, eclipsing binary]{thompson2010}, \citet{woodley2012} (purple, WD SEDs), and \citet[pink, dynamical]{watkins2015b}. Our distance is considerably smaller than previous estimates.
    }
    \label{fig:104res}
\end{figure}

For NGC\,104, we find parallax $\varpi = 0.531 \pm 0.210$~mas, RA PM $\mua = 5.50 \pm 0.70$~mas/yr, and Dec PM $\mud = -3.99 \pm 0.55$~mas/yr, with correlation terms $\rho(\varpi,\mua) = -0.440$, $\rho(\varpi,\mud) = -0.503$, and $\rho(\mua,\mud) = -0.332$. In \autoref{fig:104res}, the upper and middle panels show the parallaxes and PMs of the five TGAS stars (black) and our averages (red).

There are five previous PM estimates for NGC\,104: \citet{cudworth1993} measured $\mua = 3.4 \pm 1.7$~mas/yr and $\mud = -1.9 \pm 1.5$~mas/yr from the ground; \citet{odenkirchen1997} measured $\mua = 7.0 \pm 1.0$~mas/yr and $\mud = -5.3 \pm 1.0$~mas/yr using \textit{Hipparcos} data; \citet{freire2003} measured $\mua = 5.3 \pm 0.6$~mas/yr and $\mud = -3.3 \pm 0.6$~mas/yr using radio observations of millisecond pulsars (MSPs); \citet{anderson2003} measured $\mua = 5.88 \pm 0.18$~mas/yr and $\mud = -2.53 \pm 0.18$~mas/yr using \textit{HST}/WFPC2\footnote{We did not use the absolute PM given in \citet{anderson2003}; instead we used their measured motion of NGC\,104 relative to the Small Magellanic Cloud (SMC) combined with the SMC PM measurement from \citet{kallivayalil2006}.}; and \citet{cioni2016} measured $\mua = 7.26 \pm 0.03$~mas/yr and $\mud = -1.25 \pm 0.03$~mas/yr using ground-based data from VISTA. These points are shown in the PM panel (centre left) of \autoref{fig:104res} as orange, green, cyan, blue, and purple points, respectively.

Our PM estimate shows best agreement with the estimate derived from MSP timing by \citet{freire2003} and the \textit{HST} measurement by \citet{anderson2003}; for the latter, our RA PM is in very good agreement, though are Dec PM is slightly larger than theirs.

Our parallax of $\varpi = 0.531 \pm 0.210$~mas corresponds to a distance of $D = 1.88^{+1.23}_{-0.53}$~kpc. NGC\,104 is one of the best studied Galactic GCs and so there are many estimates of its distance in the literature; for a complete overview of previous distance determinations, see Table~2 of \citet{woodley2012}. Here we compare with only a subset of these chosen to cover a wide variety of estimation methods: \citet{gratton2003} estimated $D = 4.74^{+0.18}_{-0.17}$~kpc via MS fitting; \citet{salaris2007} estimated $D = 4.33 \pm 0.10$~kpc via HB fitting; \citet{bono2008} estimated $D = 4.61^{+0.20}_{-0.19}$~kpc using the tip of the Red Giant Branch (TRGB) and $D = 4.94^{+0.26}_{-0.24}$~kpc using RR Lyraes; \citet{thompson2010} estimated $D = 4.43^{+0.17}_{-0.16}$~kpc using an eclipsing binary; \citet{woodley2012} estimated $D = 4.69^{+0.14}_{-0.13}$~kpc using white dwarf (WD) spectral energy distributions (SEDs); and \citet{watkins2015b} used estimated $D = 4.15 \pm 0.08$~kpc using stellar kinematics. We show these distance estimates along with our own in the lower panel of \autoref{fig:104res}. Our distance estimate is considerably smaller than all of the previous distance estimates for NGC\,104.

NGC\,104 is the only cluster of the five in our sample for which the TGAS distance estimate is not consistent at 1$\sigma$ with literature estimates. It is is possible that this reflects some level of foreground contamination in our sample of five stars.\footnote{Though, as noted above, we find a $p \gg 0.05$ for the cluster, so it is statistical unlikely that any interlopers remain.} Three of the stars have individual parallax estimates consistent with the known NGC 104 distance, while the other two do not (see \autoref{fig:104res}, top-left and middle-right panels).

\subsection{NGC\,5272 (M\,3)}
\label{ssect:ppm5272}

\begin{figure}
    \centering
    \includegraphics[width=\linewidth]{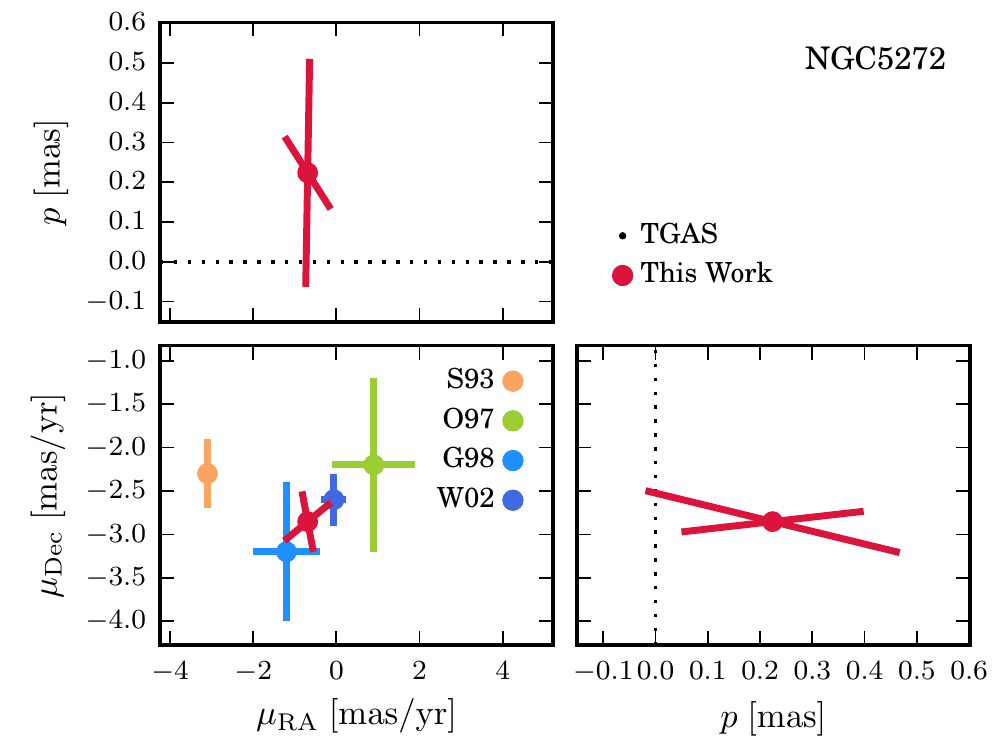}
    \includegraphics[width=\linewidth]{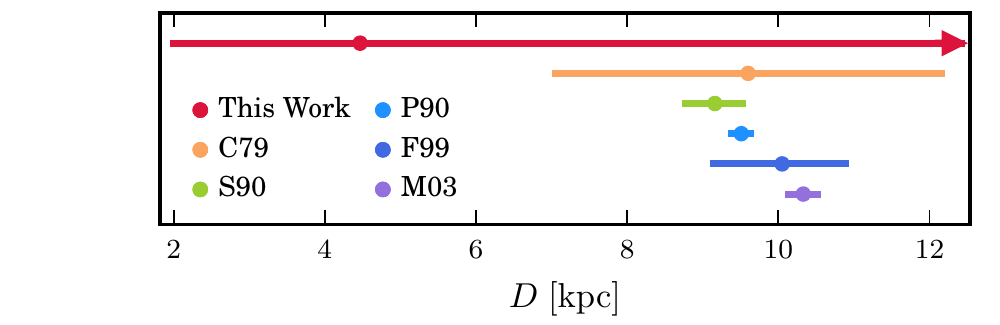}
    \caption{Upper and middle panels: parallax and PM results for NGC\,5272. The black points show the TGAS star that we identified as a cluster member. In the PM plot (middle left) we compare our estimate to values from \citet[orange, photographic plates]{scholz1993}, \citet[green, \textit{Hipparcos}]{odenkirchen1997}, \citet[cyan, \textit{Hipparcos}]{geffert1998}, and \citet[blue, photographic plates]{wu2002}. Our PM estimate is consistent with both of the latter two estimates. (There is a further estimate given given by \citet{cudworth1993}, however, this is not shown as it does not lie within the plot limits.) Lower panel: distance comparison for NGC\,5272. We invert our parallax measurement to obtain a distance estimate (red). We also show previous distance estimates from \citet[orange, dynamical]{cudworth1979}, \citet[green, MS fitting]{sandage1990}, \citet[cyan, CMD fitting]{paez1990}, \citet[blue, HB fitting]{ferraro1999}, and \citet[purple, RR Lyrae pulsations]{marconi2003}. Our distance is considerably smaller than previous estimates.
    }
    \label{fig:5272res}
\end{figure}

For NGC\,5272, we have only one member star, so we do not perform a Monte-Carlo sampling but instead report the TGAS values for the single member star. The TGAS estimate finds parallax $\varpi = 0.225 \pm 0.289$~mas, RA PM $\mua = -0.69 \pm 0.51$~mas/yr, and Dec PM $\mud = -2.85 \pm 0.37$~mas/yr, with correlation terms $\rho(\varpi,\mua) = -0.197$, $\rho(\varpi,\mud) = -0.593$, and $\rho(\mua,\mud) = 0.319$. In \autoref{fig:5272res}, the upper and middle panels show the parallax and PMs of the identified TGAS star.

There are five previous PM estimates for NGC\,5272; unfortunately there are no \textit{HST} measurements for comparison, all are ground-based estimates and they are as follows. \citet{cudworth1993} measured $\mua = -1.2 \pm 2.5$~mas/yr and $\mud = 2.4 \pm 3.0$~mas/yr from the ground; \citet{scholz1993} measured $\mua = -3.1 \pm 0.2$~mas/yr and $\mud = -2.3 \pm 0.4$~mas/yr using photographic plates; \citet{odenkirchen1997} measured $\mua = 0.9 \pm 1.0$~mas/yr and $\mud = -2.2 \pm 1.0$~mas/yr using \textit{Hipparcos} data; \citet{geffert1998} measured $\mua = -1.2 \pm 0.8$~mas/yr and $\mud = -3.2 \pm 0.8$~mas/yr also using \textit{Hipparcos} stars; and \citet{wu2002} measured $\mua = -0.06 \pm 0.3$~mas/yr and $\mud = -2.6 \pm 0.3$~mas/yr using photographic plates spanning a 70 year baseline. The latter four points are shown in the PM panel (centre left) of \autoref{fig:5272res} as orange, green, cyan, and blue points, respectively. The estimate by \citet{cudworth1993} is not shown as it is significantly offset from the others and is not located within the plot limits. The TGAS PM estimate shows best agreement with the estimate derived using photographic plates by \citet{wu2002} and the \textit{Hipparcos} measurement by \citet{geffert1998}.

Our parallax of $\varpi = 0.225 \pm 0.289$~mas corresponds to a distance of $D = 4.44^{+\infty}_{-2.50}$~kpc. The upper error bar on the distance is unconstrained, since the lower error bar on the parallax extends below zero, which, as previously discussed, is unphysical but statistically rigourous. NGC\,5272 has a number of previous estimates of its distance in the literature; here we compare with only a subset of these chosen to cover a wide variety of estimation methods: \citet{cudworth1979} estimated $D = 9.6 \pm 2.6$~kpc from cluster kinematics; \citet{sandage1990} estimated $D = 9.16^{+0.43}_{-0.41}$~kpc via MS fitting; \citet{paez1990} estimated $D = 9.51^{+0.18}_{-0.17}$~kpc using colour-magnitude diagram (CMD) fitting; \citet{ferraro1999} estimated $D = 10.05^{+0.97}_{-0.88}$~kpc using HB fitting; and \citet{marconi2003} estimated $D = 10.33 \pm 0.24$~kpc using RR Lyrae pulsations. We show these distance estimates along with our own in the lower panel of \autoref{fig:5272res}. Our distance estimate is considerably smaller than all of the previous distance estimates for NGC\,5272.

\subsection{NGC\,6121 (M\,4)}
\label{ssect:ppm6121}

\begin{figure}
    \centering
    \includegraphics[width=\linewidth]{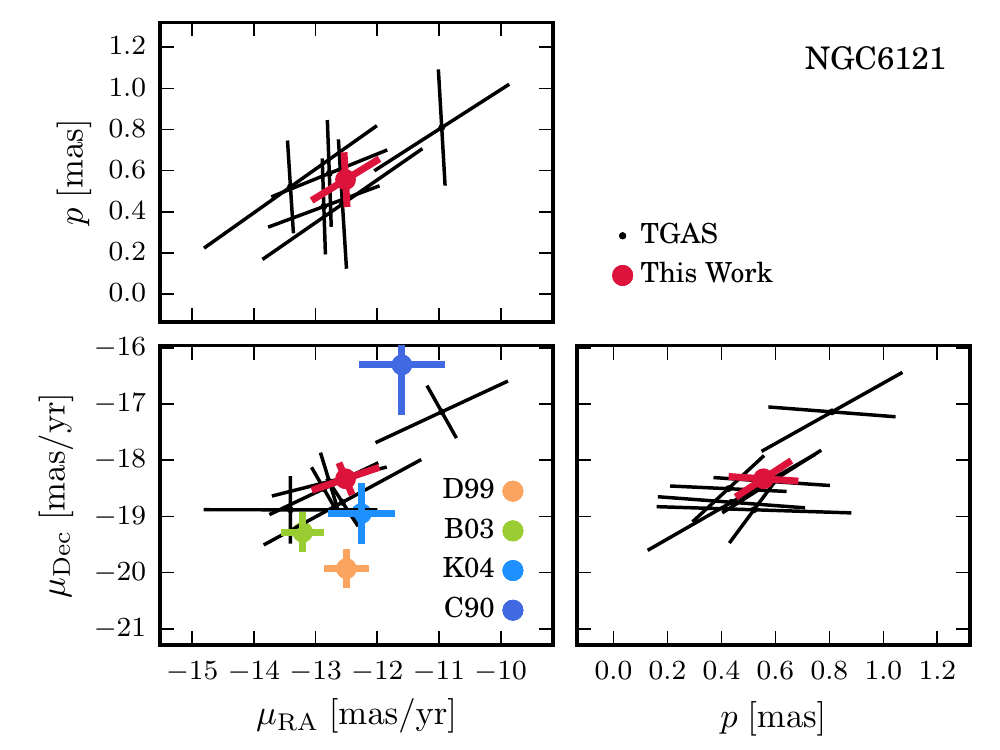}
    \includegraphics[width=\linewidth]{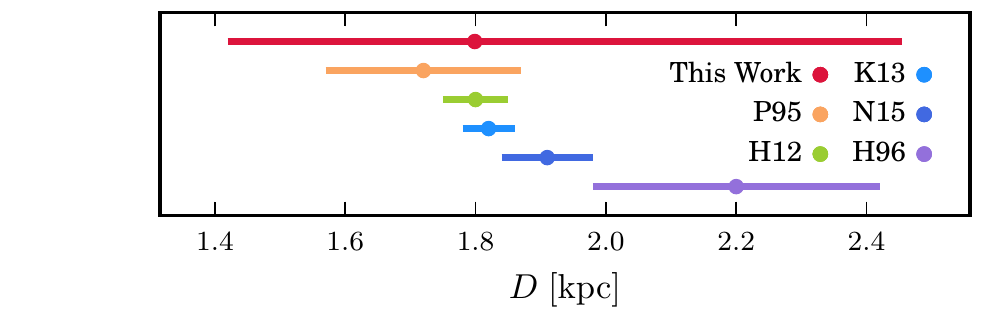}
    \caption{Upper and middle panels: parallax and PM results for NGC\,6121. The black points show the five TGAS stars that we identified as cluster members. The red points show the mean values we have calculated here. In the PM plot (middle left) we compare our estimate to values from \citet[orange, ground-based]{dinescu1999b}, \citet[green, \textit{HST}]{bedin2003}, \citet[cyan, \textit{HST}]{kalirai2004}, and \citet[blue, ground-based]{cudworth1990}. The PMs we measure are consistent with the previous \textit{HST} measurements and have similar uncertainties. Lower panel: distance comparison for NGC\,6121. We invert our parallax measurement to obtain a distance estimate (red). We also show previous distance estimates from \citet[orange, dynamical]{peterson1995}, \citet[green, HB luminosity]{hendricks2012}, \citet[cyan, eclipsing binaries]{kaluzny2013}, \citet[blue, RR Lyraes]{neeley2015}, and the estimate from \citetalias{harris1996} (purple). Our distance is consistent with previous estimates within our uncertainties, but our uncertainties are not yet competitive with existing results.}
    \label{fig:6121res}
\end{figure}

For NGC\,6121, we find parallax $\varpi = 0.556 \pm 0.149$~mas, RA PM $\mua = -12.51 \pm 0.50$~mas/yr, and Dec PM $\mud = -18.33 \pm 0.29$~mas/yr, with correlation terms $\rho(\varpi,\mua) = 0.589$, $\rho(\varpi,\mud) = 0.519$, and $\rho(\mua,\mud) = 0.488$. In \autoref{fig:6121res}, the upper and middle panels show the parallaxes and PMs of the 5 TGAS stars (black) and our averages (red).

To define membership probabilities in \autoref{ssect:probs}, we used the PM from \citetalias{casetti2013} as a reference. In fact, the PM for NGC\,6121 given in that catalogue is a combination of three separate studies and so we compare our PM again each of those studies here: \citet{dinescu1999b} measured $\mua = -12.50 \pm 0.36$~mas/yr and $\mud = -19.93 \pm 0.35$~mas/yr from SPM photographic plates using \textit{Hipparcos} stars as a reference frame; \citet{bedin2003} measured $\mua = -13.21 \pm 0.35$~mas/yr and $\mud = -19.28 \pm 0.35$~mas/yr using \textit{HST}/WFPC2 data; and \citet{kalirai2004} measured $\mua = -12.26 \pm 0.54$~mas/yr and $\mud = -18.95 \pm 0.54$~mas/yr also using \textit{HST}/WFPC2 but with a longer baseline. These points are shown in the PM panel (centre left) of \autoref{fig:6121res} as orange, green, and cyan points, respectively.

Our PM in RA is consistent with all three studies, but our PM in Dec is in much better agreement with the two \textit{HST} PMs than the ground-based PM. \textit{HST} tends to be less susceptible to various kinds of errors than ground-based measurements, so the good agreement between the \textit{HST} and TGAS estimates is very encouraging. This is also consistent with the conclusions from \citet{vandermarel2016} who recently studied rotation in the Large and Small Magellanic Clouds using TGAS and found excellent agreement between the TGAS estimates and earlier \textit{HST} studies.

There is one further estimate of the PM of NGC\,6121 that used bright field stars to create a reference frame to measure $\mua = -11.6 \pm 0.7$~mas/yr and $\mud = -16.3 \pm 0.9$~mas/yr \citep[shown in blue]{cudworth1990}, which is significantly offset from the other previous studies and our own TGAS estimate.

Our parallax of $\varpi = 0.556 \pm 0.149$~mas corresponds to a distance of $D = 1.80^{+0.66}_{-0.38}$~kpc. This is in good agreement with the distance of $D = 2.2$~kpc from \citetalias{harris1996}, which we inverted to obtain a parallax for the membership probability analysis in \autoref{ssect:probs}. As the closest GC, M4 has been well studied, so there are many estimates of its distance in the literature, and in fact the \citetalias{harris1996} value is among the largest. We summarise a selection here covering a wide variety of estimation methods: \citet{peterson1995} used PMs and LOS velocity dispersions to estimate a distance of $D = 1.72 \pm 0.14$~kpc; \citet{hendricks2012} used the luminosity of the HB to measure $D = 1.80 \pm 0.05$~kpc; \citet{kaluzny2013} used three eclipsing binaries in NGC\,6121 to measure $D = 1.82 \pm 0.04$~kpc; and \citet{neeley2015} used RR Lyraes to measure $D = 1.91 \pm 0.07$~kpc. We show these distance estimates along with our own in the lower panel of \autoref{fig:6121res}.

Our distance estimate is consistent with all previous estimates within their uncertainties, however, our distance uncertainty is considerably larger than the previous results.

\subsection{NGC\,6397}
\label{ssect:ppm6397}

\begin{figure}
    \centering
    \includegraphics[width=\linewidth]{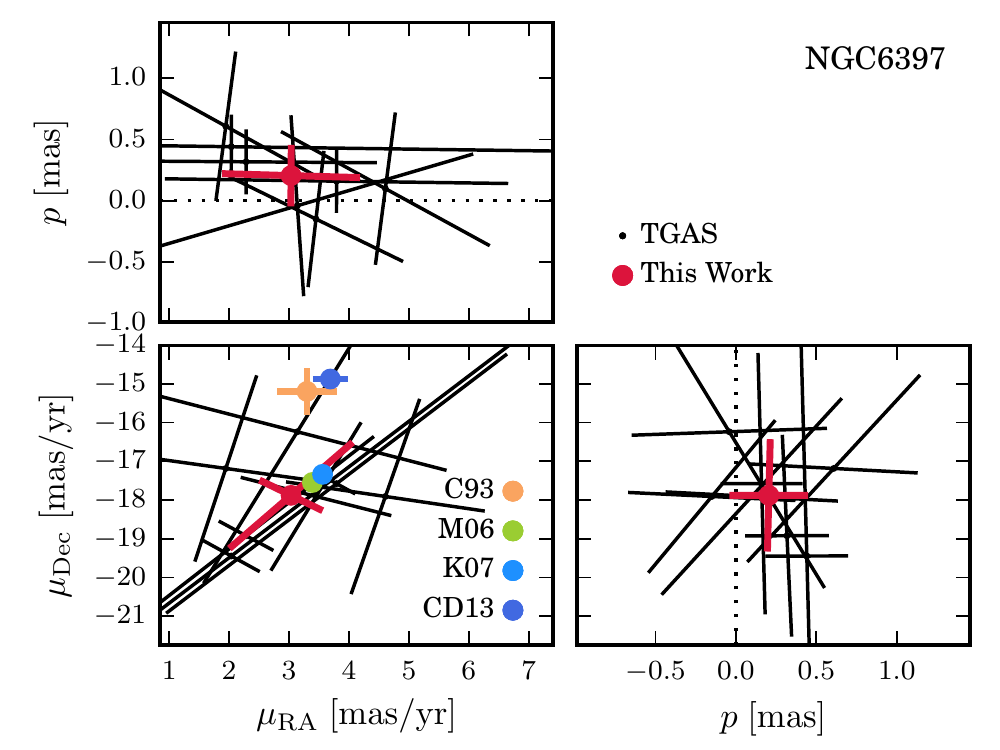}
    \includegraphics[width=\linewidth]{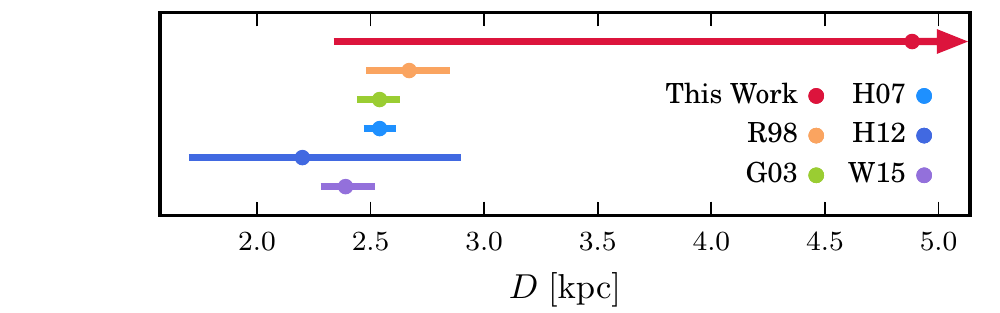}
    \caption{Upper and middle panels: parallax and PM results for NGC\,6397. The black points show the seven TGAS stars that we identified as cluster members. The red points show the mean values we have calculated here. In the PM plot (middle left) we compare our estimate to values from \citet[orange, ground-based]{cudworth1993}, \citet[green, \textit{HST}]{milone2006}, \citet[cyan, \textit{HST}]{kalirai2007}, and \citetalias{casetti2013} (blue, ground-based). The PMs we measure are consistent with the previous \textit{HST} measurements, albeit with uncertainties an order of magnitude larger and significantly offset from the previous ground-based measurements. Lower panel: distance comparison for NGC\,6397. We invert our parallax measurement to obtain a distance estimate (red); no upper error bar is shown, since lower error bar on the parallax extended below zero, corresponding to unphysical results but effectively indicating that the upper limit on the distance uncertainty is infinite. We also show previous distance estimates from \citet[orange, MS fitting]{reid1998}, \citet[green, MS fitting]{gratton2003}, \citet[cyan, WD sequence]{hansen2007}, \citet[blue, dynamical]{heyl2012}, and the estimate from \citet[purple, dynamical]{watkins2015b}. Our distance uncertainties are significantly larger than previous estimates, however, our distance estimate is consistent within the previous results within the very large error bars.}
    \label{fig:6397res}
\end{figure}

For NGC\,6397, we find parallax $\varpi = 0.205 \pm 0.223$~mas, RA PM $\mua = 3.03 \pm 1.09$~mas/yr, and Dec PM $\mud = -17.88 \pm 1.36$~mas/yr, with correlation terms $\rho(\varpi,\mua) = -0.066$, $\rho(\varpi,\mud) = 0.036$, and $\rho(\mua,\mud) = 0.755$. In \autoref{fig:6397res}, the upper and middle panels show the parallaxes and PMs of the seven TGAS stars (black) and our averages (red).

There are four previous PM estimates for NGC\,6397: \citet{cudworth1993} measured $\mua = 3.3 \pm 0.5$~mas/yr and $\mud = -15.2 \pm 0.6$~mas/yr from the ground; \citet{milone2006} measured $\mua = 3.39 \pm 0.15$~mas/yr and $\mud = -17.55 \pm 0.15$~mas/yr using \textit{HST}/WFPC2 data; \citet{kalirai2007} measured $\mua = 3.56 \pm 0.04$~mas/yr and $\mud = -17.34 \pm 0.04$~mas/yr using a combination of \textit{HST}/WFPC2 and \textit{HST}/ACS data; and \citetalias{casetti2013} measured $\mua = 3.69 \pm 0.29$~mas/yr and $\mud = -14.88 \pm 0.26$~mas/yr using a combination of photographic plates and ground-based CCD images. These points are shown in the PM panel (centre left) of \autoref{fig:6397res} as orange, green, cyan, and blue points, respectively. Note that all of these measurements are in reasonable agreement in $\mua$, but are discrepant at $\sim 2$~mas/yr in $\mud$. We find that our $\mud$ estimate is in much better agreement with the two \textit{HST} PMs than the two ground-based PMs. This trend is consistent with our findings for NGC\,6121 in \autoref{ssect:ppm6121}.

Our parallax of $\varpi = 0.205 \pm 0.223$~mas corresponds to a distance of $D = 4.88^{+\infty}_{-2.55}$~kpc. Our upper error bar on the distance is unconstrained, since the lower error bar on the parallax for NGC\,6397 extends below zero, which, as previously discussed, is unphysical but statistically rigourous.

There are a number of previous distance estimates in the literature, of which we show a representative sample here, obtained using different methods: \citet{reid1998} estimated $D = 2.67^{+0.19}_{-0.18}$~kpc and \citet{gratton2003} estimated $D = 2.54^{+0.10}_{-0.09}$~kpc, both via MS fitting; \citet{hansen2007} measured $D = 2.54 \pm 0.07$~kpc by fitting the WD sequence; and \citet{heyl2012} estimated $D = 2.2^{+0.5}_{-0.7}$~kpc and \citet{watkins2015b} estimated $D = 2.39^{+0.13}_{-0.11}$~kpc by combining PMs and LOS velocities. We show these distance estimates along with our own in the lower panel of \autoref{fig:6397res}. Our result is larger than all of these previous estimates, but is consistent within $1 \sigma$.

\subsection{NGC\,6656 (M\,22)}
\label{ssect:ppm6656}

\begin{figure}
    \centering
    \includegraphics[width=\linewidth]{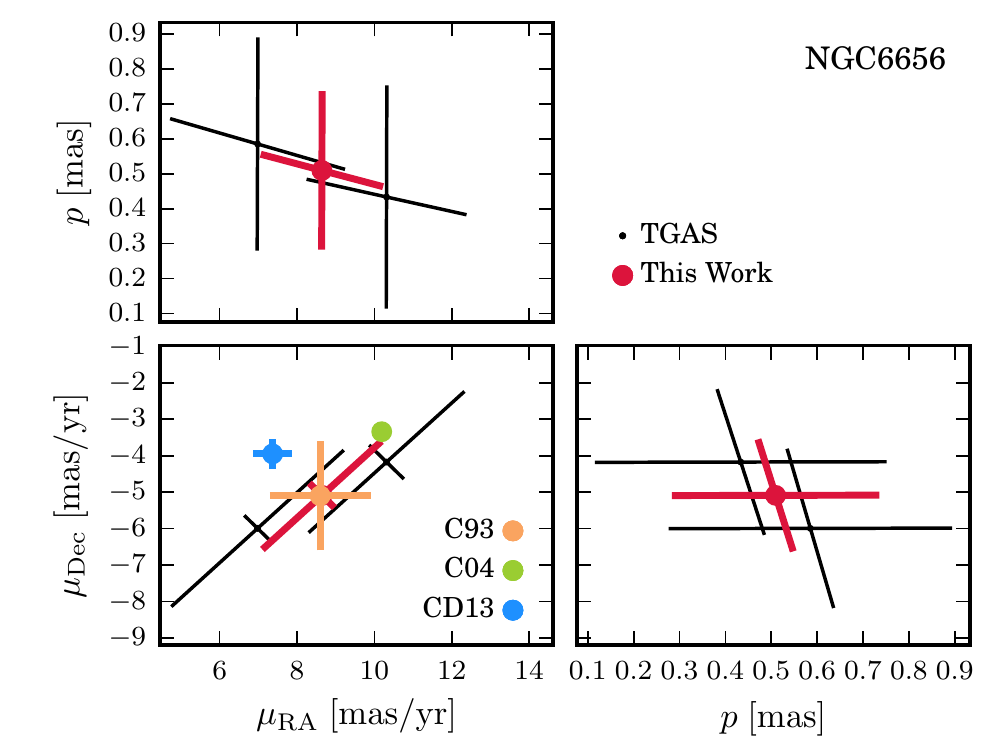}
    \includegraphics[width=\linewidth]{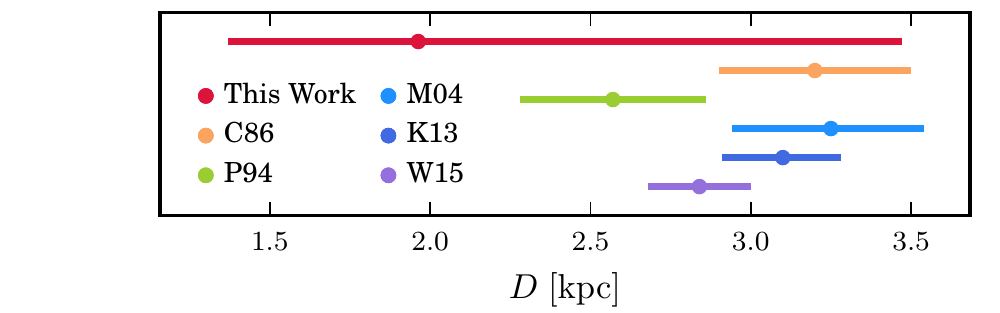}
    \caption{Upper and middle panels: parallax and PM results for NGC\,6656. The black points show the two TGAS stars that we identified as cluster members. The red points show the mean values we have calculated here. In the PM plot (middle left) we compare our estimate to values from \citet[orange, ground-based]{cudworth1993}, \citet[green, \textit{HST}]{chen2004}, and \citetalias{casetti2013} (cyan, ground-based). The PMs we measure are consistent with the previous measurements, albeit with uncertainties an order of magnitude larger. Lower panel: distance comparison for NGC\,6656. We invert our parallax measurement to obtain a distance estimate (red). We also show previous distance estimates from \citet[orange, HB fitting]{cudworth1986}, \citet[green, dynamical]{peterson1994}, \citet[cyan, HB fitting]{monaco2004}, \citet[blue, RR Lyraes]{kunder2013}, and \citet[purple, dynamical]{watkins2015b}. Our distance uncertainties are significantly larger than previous estimates, however, our distance estimate is consistent within the previous results within the our uncertainties.}
    \label{fig:6656res}
\end{figure}

For NGC\,6656, we find parallax $\varpi = 0.509 \pm 0.222$~mas, RA PM $\mua = 8.64 \pm 1.49$~mas/yr, and Dec PM $\mud = -5.09 \pm 1.45$~mas/yr, with correlation terms $\rho(\varpi,\mua) = -0.194$, $\rho(\varpi,\mud) = -0.160$, and $\rho(\mua,\mud) = 0.935$. In \autoref{fig:6656res}, the upper and middle panels show the parallaxes and PMs of the two TGAS stars (black) and our averages (red).

There are three previous PM estimates for NGC\,6656: \citet{cudworth1993} measured $\mua = 8.6 \pm 1.3$~mas/yr and $\mud = -5.1 \pm 1.5$~mas/yr from the ground; \citet{chen2004} measured $\mua = 10.19 \pm 0.20$~mas/yr and $\mud = -3.34 \pm 0.10$~mas/yr using \textit{HST}/WFPC2 data; and \citetalias{casetti2013} measured $\mua = 7.37 \pm 0.50$~mas/yr and $\mud = -3.95 \pm 0.42$~mas/yr using a combination of photographic plates and ground-based CCD images. These points are shown in the PM panel (centre left) of \autoref{fig:6656res} as orange, green, and cyan points, respectively. \citetalias{casetti2013} advises caution considering the \citet{chen2004} \textit{HST} result, as they note that bulge stars were used to create the reference frame and that the possible (and likely) motion of the bulge stars was not factored into the analysis, nevertheless we do make a comparison here.

We find that our PM estimate is consistent with all three studies, indeed it is in exceptional agreement with the ground-based PM estimate from \citet{cudworth1993}. Contrary to the two previous clusters, we do not find the best agreement with the previous \textit{HST} estimate, possibly for the reasons discussed above.

Our parallax of $\varpi = 0.509 \pm 0.222$~mas corresponds to a distance of $D = 1.96^{+1.51}_{-0.59}$~kpc. There are a number of previous distance estimates in the literature obtained using different methods: \citet{cudworth1986} estimated $D = 3.2 \pm 0.3$~kpc \citepalias[which is the value quoted in][]{harris1996} and \citet{monaco2004} estimated $D = 3.25^{+0.31}_{-0.29}$~kpc, both via HB fitting; \citet{peterson1994} measured $D = 2.57 \pm 0.29$~kpc and \citet{watkins2015b} estimated $D = 2.84 \pm 0.16$~kpc by combining PMs and LOS velocities; and \citet{kunder2013} estimated $D = 3.10^{+0.19}_{-0.18}$~kpc using RR Lyraes. We show these distance estimates along with our own in the lower panel of \autoref{fig:6656res}. This time, our result is lower than all of the previous estimates, but is consistent within the error bars. Once again, our distance uncertainties are an order of magnitude larger than for previous estimates.


\section{Absolute Space Motions}
\label{sect:spacemotion}

We can combine our PM estimates with sky coordinates, distances, and heliocentric RVs to determine the implied Galactocentric motion of the GCs. We use our PM estimates for this analysis, but not our distance (parallax) estimates, as it is clear those are less precise than literature values. The sky coordinates and distances we used here are listed in \autoref{table:positions}.

\begin{table}
    \caption{Additional cluster properties for space motion analysis.}
    \label{table:positions}
    \centering
    
    \begin{tabular}{cccc}
        \hline
        \hline
        Cluster & $\alpha$ & $\delta$ & $D$ \\
        & (deg) & (deg) & (kpc) \\
        (1) & (2) & (3) & (4) \\
        \hline \\[-1em]
        NGC\,104  &   6.024 & -72.081 & $4.61^{+0.20}_{-0.19}$ \\[0.4em]
        NGC\,5272 & 205.548 &  28.377 & $10.05^{+0.97}_{-0.88}$ \\[0.4em]
        NGC\,6121 & 245.897 & -26.526 & 1.80 $\pm$ 0.05 \\[0.4em]
        NGC\,6397 & 265.175 & -53.674 & $2.54^{+0.10}_{-0.09}$ \\[0.4em]
        NGC\,6656 & 279.100 & -23.905 & $3.10^{+0.19}_{-0.18}$ \\[0.4em]
        \hline
    \end{tabular}
    
    \qquad
    
    \textbf{Notes.} Columns: (1) cluster name; (2) Right Ascension; (3) Declination; (4) adopted distance (sources: \citet[][NGC\,104]{bono2008}, \citet[][NGC\,5272]{ferraro1999}, \citet[][NGC\,6121]{hendricks2012}, \citet[][NGC\,6397]{gratton2003}, and \citet[][NGC\,6656]{kunder2013}).
    
\end{table}

We adopt a Cartesian coordinate system $(X,Y,Z)$ centred on the Galactic Centre, where the $X$-axis points in the direction from the Sun to the Galactic Centre, the $Y$-axis points in the direction of the Sun's Galactic rotation, and the $Z$-axis points toward the North Galactic Pole.

To transform the measured heliocentric velocities into velocities $(U,V,W)$ in the Galactocentric rest frame, we adopt a distance from the Sun to the Galactic Centre of $R_0 = 8.29 \pm 0.16$~kpc, and a circular velocity of the local standard of rest (LSR) of $V_0 = 239 \pm 5$~km/s \citep[both][]{mcmillan2011}. We also assume a solar peculiar velocity relative to the LSR of $(U_{\rm pec}, V_{\rm pec}, W_{\rm pec}) = (11.10 \pm 1.23, 12.24 \pm 2.05, 7.25 \pm 0.63)$~km/s \citep{schoenrich2010}.

To estimate Galactocentric velocities and their uncertainties, we use a Monte-Carlo scheme that propagates all observational distance and velocity uncertainties and their correlations, including those for the Sun. We also provide Galactocentric radial $\vrad$ and tangential $\vtan$ velocities defined in a spherical coordinate system, where $\vrad$ is positive outward from the Galactic Centre and $\vtan$ is the magnitude of the motion perpendicular to the radial motion. We also calculate the radial $\Pi$ and tangential $\Theta$ velocities in a cylindrical coordinate system, where $\Pi$ is again positive outward from the Galactic Centre and $\Theta$ is positive in the direction of Galactic rotation. Our results are summarised in \autoref{table:orbits}.

\begin{table*}
    \caption{Cluster space motions and orbits.}
    \label{table:orbits}
    \centering
    \setlength{\tabcolsep}{1pt}
    
    \begin{tabular}{cccccccccccc}
        \hline
        \hline
        Cluster & $X$ & $Y$ & $Z$ & $U$ & $V$ & $W$ & $|\mathbf{v}|$ & $v_{\rm rad}$ & $v_{\rm tan}$ & $\Pi$ & $\Theta$ \\
        & (kpc) & (kpc) & (kpc) & (km/s) & (km/s) & (km/s) & (km/s) & (km/s) & (km/s) & (km/s) & (km/s) \\
        (1) & (2) & (3) & (4) & (5) & (6) & (7) & (8) & (9) & (10) & (11) & (12) \\
        \hline \\[-1em]
        NGC\,104 & -6.38 & -2.65 & -3.25 & -74.97 $\pm$ 14.10 & 145.23 $\pm$ 13.12 &  68.30 $\pm$ 8.85 & 177.14 $\pm$ 6.81 & -16.84 $\pm$ 14.81 & 176.34 $\pm$ 7.02 &  13.54 $\pm$ 15.20 & 163.03 $\pm$ 10.51 \\[0.4em]
        NGC\,5272 & -6.83 &  1.32 &  9.86 &  49.87 $\pm$ 22.34 & 106.00 $\pm$ 23.90 & -129.51 $\pm$ 4.83 & 174.63 $\pm$ 19.44 & -122.42 $\pm$ 15.71 & 124.54 $\pm$ 23.87 & -28.71 $\pm$ 21.15 & 113.36 $\pm$ 24.03 \\[0.4em]
        NGC\,6121 & -6.58 & -0.27 &  0.50 &  55.10 $\pm$ 2.02 &  50.19 $\pm$ 8.42 &   2.13 $\pm$ 3.97 &  74.57 $\pm$ 6.38 & -56.81 $\pm$ 2.35 &  48.30 $\pm$ 8.31 & -57.07 $\pm$ 2.16 &  47.87 $\pm$ 8.47 \\[0.4em]
        NGC\,6397 & -5.98 & -0.92 & -0.53 & -61.20 $\pm$ 7.73 &  98.79 $\pm$ 16.28 & -132.62 $\pm$ 14.60 & 176.34 $\pm$ 12.75 &  56.72 $\pm$ 10.39 & 166.97 $\pm$ 13.49 &  45.52 $\pm$ 9.48 & 106.88 $\pm$ 14.93 \\[0.4em]
        NGC\,6656 & -5.26 &  0.53 & -0.41 & -148.98 $\pm$ 4.97 & 212.58 $\pm$ 21.89 & -119.23 $\pm$ 23.26 & 285.66 $\pm$ 21.25 & 178.12 $\pm$ 8.62 & 223.33 $\pm$ 21.22 & 169.28 $\pm$ 7.19 & 196.49 $\pm$ 21.69 \\[0.4em]
        \hline
    \end{tabular}

    \qquad
    
    \textbf{Notes.} Columns: (1) cluster name; (2) $X$ position coordinate; (3) $Y$ position coordinate; (4) $Z$ position coordinate; (5) $U$ velocity component; (6) $V$ velocity component; (7) $W$ velocity component; (8) magnitude of total velocity; (9) radial component of velocity in a spherical coordinate system; (10) tangential component of velocity in a spherical coordinate system; (11) radial component of velocity in a cylindrical coordinate system; (12) tangential component of velocity in a cylindrical coordinate system.
    
\end{table*}

\citet{zinn1985} separated Galactic GCs into two subpopulations based on their metallicities and orbital properties: those belonging to the disk ($\feh > -0.8$) and those belonging to the halo ($\feh < -0.8$). Overall, the disk clusters have a lower orbital eccentricity and are confined to the Galactic plane, while the halo clusters span a range of eccentricities and explore greater distances from the disk plane.

In their orbital analysis, \citet{odenkirchen1997} noted that the orbital energy of \textbf{NGC\,104} is very close to its corresponding circular orbit (ie.\ the energy of a circular orbit at its current Galactocentric distance). They noted that the orbital properties of NGC\,104 were somewhat offset from the rest of their sample, which were all thought to be halo clusters, and so concluded that NGC\,104 is a disk cluster. This conclusion was also reached by \citet{cudworth1993}, who noted that the space velocity of NGC\,104 marks it as a thick-disk cluster, despite its relatively large distance from the plane.

\textbf{NGC\,5272} is currently located almost 10~kpc above the plane of the disk, so it is extremely likely to be a halo cluster. All five of the previous PM studies discussed in \autoref{ssect:ppm5272} performed space motion analyses. \citet{cudworth1993} found that the orbit of NGC\,5272 was very uncertain in their analysis; nevertheless, they concluded that the orbit was more consistent with the \citet{zinn1985} halo population, as expected. Their space motion was somewhat different from the other studies, but all of the other studies reached broadly similar conclusions: NGC\,5272 is a halo cluster with a box-like orbit \citep{scholz1993,wu2002} and is likely near its apocentre \citep{geffert1998}.

\textbf{NGC\,6121} is located very close to the Galactic plane and has a very small $Z$ component of velocity, which implies that it spends most of its time in the disk. However, both \citet{dinescu1999b} and \citet{bedin2003} found very radial orbits for NGC\,6121, and their extreme eccentricities favour NGC\,6121 as being a halo cluster on a very low inclination orbit and not a disk cluster. The PM determination of \citet{cudworth1993} was significantly offset from all other previous estimates, however, their conclusions on the orbit of NGC\,6121 were nonetheless broadly similar.

Our space motion for NGC\,6121 is in reasonable agreement with the analyses by \citet{dinescu1999b} and \citet{bedin2003}, but shows best agreement with that calculated by \citet{kalirai2004} -- which is to be expected given the very good agreement between our PM determination and theirs -- except for the sign of $W$. This sign discrepancy is not surprising, as we find a very small $W$ component that is consistent with both positive and negative $W$ velocities within 1$\sigma$. Small changes in the assumed parameters will change the sign of the $W$ velocity, but this has little effect on our overall conclusions.

\citet{dinescu1999b} also estimated the tidal-shock rate of NGC\,6121 based on its orbit and found that this is larger than the two-body relaxation time in the cluster; this indicates that tidal shocks due to the bulge and the disk have likely had a significant influence on the evolution of the cluster. Again, it is not surprising that a cluster confined to the disk should be strongly affected by tides.

\textbf{NGC\,6397} is also currently located close to the disk plane, however, the high W velocity implies that it does move away from the plane and is most likely a halo cluster, as expected from its low $\feh$ metallicity. This is consistent with more detailed orbital analyses by \citet{cudworth1993} and \citetalias{casetti2013} which concluded that NGC\,6397 has a ``typical halo-like orbit'' \citep[following][]{zinn1985}. \citet{milone2006} found that NGC\,6397 has a boxy orbit that oscillates rapidly through the disk. Thanks to the moderate eccentricity of the orbit, NGC\,6397 does not move too far from the Galactic centre; thus, it passes through high-density regions of the disk and is expected to be affected by tidal shocks.

Similarly, \textbf{NGC\,6656} is also currently located close to the disk plane but its $W$ velocity suggests that it does move away from the disk, so it is more likely to be a halo cluster. From detailed studies of the implied orbit, both \citet{cudworth1993} and \citetalias{casetti2013} also concluded that NGC\,6656 is a halo cluster. This is not surprising given that NGC\,6656 is metal-poor and also shows signs of inhomogeneous element abundances \citep[e.g.,][]{norris1983}, which is thought to indicate extended star formation \citep[e.g.,][]{marino2011} and point to an extragalactic origin \citep[e.g.,][]{lee2009}. NGC\,6656 is also likely to be vulnerable to tides from repeated passages through the disk. This is borne out by observations: \citet{ness2013} found a population of stars in their ARGOS study of the bulge that had properties consistent with NGC\,6656, which could be extratidal stars from the cluster. \citet{kunder2014} also found a candidate extratidal star of NGC\,6656 in RAVE data.


\section{Discussion and Conclusions}
\label{sect:conclusions}

In this pilot study, we have identified a total of 20 members of 5 Galactic GCs in the TGAS catalogue from \textit{Gaia} DR1 -- 5 stars in NGC\,104 (47\,Tuc), 1 star in NGC\,5272 (M\,3), 5 stars in NGC\,6121 (M\,4), 7 stars in NGC\,6397, and 2 stars in NGC\,6656 -- and used these stars to determine the PMs and parallaxes of these clusters. These first Gaia PMs for GCs provide new insights into their dynamics. The parallaxes instead, while consistent with existing knowledge, are not yet competitive with other methods for determining GC distances.

\begin{figure}
    \centering
    \includegraphics[width=\linewidth]{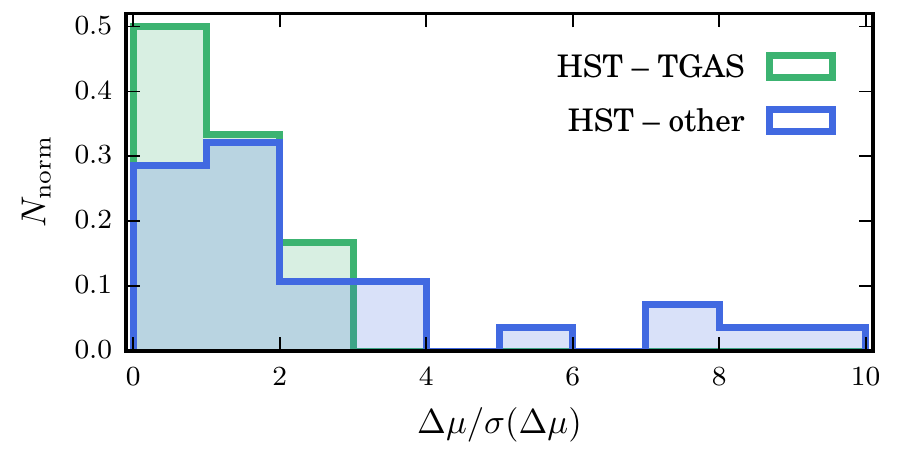}
    \caption{Normalised histograms of PM differences $\Delta \mu$ as a fraction of the uncertainty in the difference $\sigma(\Delta \mu)$ for all clusters, with RA PMs and Dec PMs considered independently. The blue histogram shows the offsets between \textit{HST} PM measurements and other non-\textit{HST} PM measurements, all taken from the literature (see \autoref{sect:ppms}). In some cases, the \textit{HST} and non-\textit{HST} measurements are discrepant at more than $5\sigma$. The green histogram shows the offsets between literature \textit{HST} PM measurements and our TGAS PM measurements. All \textit{HST} and TGAS measurements agree within $3\sigma$.}
    \label{fig:pmdiffs}
\end{figure}

We reviewed all existing PM measurements for the GCs in our final sample. Existing PM estimates for the same cluster do not generally agree with each other within the random errors; in some cases, measurements disagree by $5\sigma$ or more (see \autoref{fig:pmdiffs}, blue histogram). Hence, the accuracy of the existing PM measurements is generally dominated by systematic errors, and not random errors, as is often the case in astrometry. By contrast, the TGAS catalogue combines measurements from \textit{Hipparcos} and \textit{Gaia}: two space missions with stable and well-calibrated platforms designed specifically to do astrometry. The main value of our TGAS analysis is therefore that it provides a homogeneous study of multiple GCs based on an astrometric catalogue with well-controlled systematics.

We have not explicitly included possible \textit{spatial} correlations in TGAS PM errors \citep{gaia2016, lindegren2016} in our analysis. The effect of such correlations would be to underestimate the random error in the weighted average PM of a stellar sample \citep{kroupa1997}. \textit{Gaia} DR1 has a nominal systematic positional accuracy of $\sim 0.3$~mas. Over the $\sim 25$~yr time baseline with the \textit{Hipparcos} mission, this introduces a PM error of only $\sim 0.01$~mas/yr. \citet{vandermarel2016} found the agreement between TGAS and \textit{HST} PMs for the Magellanic Clouds to be significantly better than $\lesssim 0.1$~mas/yr. This further bounds the size of any possible PM systematics. Systematic PM errors in the TGAS PMs are therefore well below the size of the random PM errors for our GCs in \autoref{table:results} (which range from $0.29$ to $1.49$~mas/yr).

The random PM errors from our analysis are comparable to those from many existing studies, but do not yet improve upon them. In general, the PM diagrams presented here show that our measurements are consistent with the ``scatter clouds" of measurements from previous studies. A detailed assessment shows that our measurements are generally fully consistent with existing \textit{HST}-based measurements (see \autoref{fig:pmdiffs}, green histogram), but not with all existing ground-based measurements. This agrees with the findings of \citet{vandermarel2016}, who found excellent agreement between measurements of the PMs of the Magellanic Clouds from TGAS and \textit{HST}. We therefore conclude that: (1) since TGAS and \textit{HST} use very different methods for astrometry, our GC results provide further external validation of both data sets and their underlying approaches; and (2) many ground-based GC PM measurements suffer from systematic uncertainties in excess of the random errors.

We also combined our PMs with literature RVs and distances to calculate absolute Galactocentric space motions for the clusters, corrected for peculiar motion of the Sun relative to the LSR and for rotation of the LSR about the Galactic Centre. Our space motions are broadly consistent with previous orbital analyses.

Our results highlight the promise of future \textit{Gaia} data releases for the determination of parallaxes and PMs in Galactic GCs. Comparing parallax distances from \textit{Gaia} with previous photometric and dynamical distance estimates will provide crucial verification of all three methods, and improved distances and PMs will further constrain GC orbits. With future releases, \textit{Gaia} PMs will be able to resolve the internal motions of nearby Galactic GCs as well as measure global motions as we have done here.

\textit{Gaia} PMs and \textit{HST} PMs are highly complementary, as the two observatories have different strengths. At present, the majority of \textit{HST} PMs for the inner GCs are relative, due to the difficulty of finding sufficient fixed background sources in the crowded GC fields that can be used to create an absolute reference frame \citep{bellini2014}. This does not affect velocity dispersion or anisotropy determination \citep{watkins2015a} but does not allow for measurement of rotation, or indeed the global motion of the GC. \textit{Gaia} PMs are absolute, so they will provide crucial constraints on cluster rotation and on their global motions. By contrast, \textit{HST} can measure PMs down to fainter magnitudes, so it can go deeper for the nearby clusters, and can measure PMs for more distant clusters than \textit{Gaia} will be able to achieve. Thus, \textit{HST} PMs will be crucial for constraining the shape of the outer MW halo. \textit{HST} will also provide more accurate astrometry in the crowded centres of clusters (though these regions are challenging even for \textit{HST}).

For now, \textit{HST} remains the best source of GC PMs, however, in the future the the best choice will depend on the particular science goals. Combined together, \textit{Gaia} and \textit{HST} PMs will provide the greatest coverage, depth, and accuracy of all and will revolutionise our understanding of both individual clusters and the entire MW GC population.


\section*{Acknowledgements}

We thank the anonymous referee for a useful report. L.L.W.\ wishes to thank Erik Tollerud for useful discussions. This research has made use of Astropy\footnote{\url{http://www.astropy.org}}, a community-developed core Python package for Astronomy \citep{astropy2013}. This research has made use of NASA's Astrophysics Data System. This work has made use of data from the European Space Agency (ESA) space mission \textit{Gaia}\footnote{\url{http://www.cosmos.esa.int/gaia}}, processed by the \textit{Gaia} Data Processing and Analysis Consortium (DPAC)\footnote{\url{http://www.cosmos.esa.int/web/gaia/dpac/consortium}}. Funding for the DPAC has been provided by national institutions, in particular the institutions participating in the \textit{Gaia} Multilateral Agreement.



\bibliographystyle{aasjournal}

\bibliography{refs}

\begin{thebibliography}{}
\expandafter\ifx\csname natexlab\endcsname\relax\def\natexlab#1{#1}\fi
\providecommand{\url}[1]{\href{#1}{#1}}

\bibitem[{{Anderson} \& {King}(2003)}]{anderson2003}
{Anderson}, J., \& {King}, I.~R. 2003, \aj, 126, 772

\bibitem[{{Astraatmadja} \& {Bailer-Jones}(2016)}]{astraatmadja2016}
{Astraatmadja}, T.~L., \& {Bailer-Jones}, C.~A.~L. 2016, \apj, 832, 137

\bibitem[{{Astropy Collaboration} {et~al.}(2013){Astropy Collaboration},
  {Robitaille}, {Tollerud}, {Greenfield}, {Droettboom}, {Bray}, {Aldcroft},
  {Davis}, {Ginsburg}, {Price-Whelan}, {Kerzendorf}, {Conley}, {Crighton},
  {Barbary}, {Muna}, {Ferguson}, {Grollier}, {Parikh}, {Nair}, {Unther},
  {Deil}, {Woillez}, {Conseil}, {Kramer}, {Turner}, {Singer}, {Fox}, {Weaver},
  {Zabalza}, {Edwards}, {Azalee Bostroem}, {Burke}, {Casey}, {Crawford},
  {Dencheva}, {Ely}, {Jenness}, {Labrie}, {Lim}, {Pierfederici}, {Pontzen},
  {Ptak}, {Refsdal}, {Servillat}, \& {Streicher}}]{astropy2013}
{Astropy Collaboration}, {Robitaille}, T.~P., {Tollerud}, E.~J., {et~al.} 2013,
  \aap, 558, A33

\bibitem[{{Bedin} {et~al.}(2003){Bedin}, {Piotto}, {King}, \&
  {Anderson}}]{bedin2003}
{Bedin}, L.~R., {Piotto}, G., {King}, I.~R., \& {Anderson}, J. 2003, \aj, 126,
  247

\bibitem[{{Bellini} {et~al.}(2014){Bellini}, {Anderson}, {van der Marel},
  {Watkins}, {King}, {Bianchini}, {Chanam{\'e}}, {Chandar}, {Cool}, {Ferraro},
  {Ford}, \& {Massari}}]{bellini2014}
{Bellini}, A., {Anderson}, J., {van der Marel}, R.~P., {et~al.} 2014, \apj,
  797, 115

\bibitem[{{Bono} {et~al.}(2008){Bono}, {Stetson}, {Sanna}, {Piersimoni},
  {Freyhammer}, {Bouzid}, {Buonanno}, {Calamida}, {Caputo}, {Corsi}, {Di
  Cecco}, {Dall'Ora}, {Ferraro}, {Iannicola}, {Monelli}, {Nonino}, {Pulone},
  {Sterken}, {Storm}, {Tuvikene}, \& {Walker}}]{bono2008}
{Bono}, G., {Stetson}, P.~B., {Sanna}, N., {et~al.} 2008, \apjl, 686, L87

\bibitem[{{Casetti-Dinescu} {et~al.}(2007){Casetti-Dinescu}, {Girard},
  {Herrera}, {van Altena}, {L{\'o}pez}, \& {Castillo}}]{casetti2007}
{Casetti-Dinescu}, D.~I., {Girard}, T.~M., {Herrera}, D., {et~al.} 2007, \aj,
  134, 195

\bibitem[{{Casetti-Dinescu} {et~al.}(2013){Casetti-Dinescu}, {Girard},
  {J{\'{\i}}lkov{\'a}}, {van Altena}, {Podest{\'a}}, \&
  {L{\'o}pez}}]{casetti2013}
{Casetti-Dinescu}, D.~I., {Girard}, T.~M., {J{\'{\i}}lkov{\'a}}, L., {et~al.}
  2013, \aj, 146, 33

\bibitem[{{Casetti-Dinescu} {et~al.}(2010){Casetti-Dinescu}, {Girard},
  {Korchagin}, {van Altena}, \& {L{\'o}pez}}]{casetti2010}
{Casetti-Dinescu}, D.~I., {Girard}, T.~M., {Korchagin}, V.~I., {van Altena},
  W.~F., \& {L{\'o}pez}, C.~E. 2010, \aj, 140, 1282

\bibitem[{{Chaboyer}(1995)}]{chaboyer1995}
{Chaboyer}, B. 1995, \apjl, 444, L9

\bibitem[{{Chen} {et~al.}(2004){Chen}, {Chen}, \& {Wang}}]{chen2004}
{Chen}, D., {Chen}, L., \& {Wang}, J.-J. 2004, Chinese Physics Letters, 21,
  1673

\bibitem[{{Cioni} {et~al.}(2016){Cioni}, {Bekki}, {Girardi}, {de Grijs},
  {Irwin}, {Ivanov}, {Marconi}, {Oliveira}, {Piatti}, {Ripepi}, \& {van
  Loon}}]{cioni2016}
{Cioni}, M.-R.~L., {Bekki}, K., {Girardi}, L., {et~al.} 2016, \aap, 586, A77

\bibitem[{{Cudworth}(1979)}]{cudworth1979}
{Cudworth}, K.~M. 1979, \aj, 84, 1312

\bibitem[{{Cudworth}(1986)}]{cudworth1986}
---. 1986, \aj, 92, 348

\bibitem[{{Cudworth} \& {Hanson}(1993)}]{cudworth1993}
{Cudworth}, K.~M., \& {Hanson}, R.~B. 1993, \aj, 105, 168

\bibitem[{{Cudworth} \& {Rees}(1990)}]{cudworth1990}
{Cudworth}, K.~M., \& {Rees}, R. 1990, \aj, 99, 1491

\bibitem[{{Dinescu} {et~al.}(1999{\natexlab{a}}){Dinescu}, {Girard}, \& {van
  Altena}}]{dinescu1999b}
{Dinescu}, D.~I., {Girard}, T.~M., \& {van Altena}, W.~F. 1999{\natexlab{a}},
  \aj, 117, 1792

\bibitem[{{Dinescu} {et~al.}(2003){Dinescu}, {Girard}, {van Altena}, \&
  {L{\'o}pez}}]{dinescu2003}
{Dinescu}, D.~I., {Girard}, T.~M., {van Altena}, W.~F., \& {L{\'o}pez}, C.~E.
  2003, \aj, 125, 1373

\bibitem[{{Dinescu} {et~al.}(1997){Dinescu}, {Girard}, {van Altena}, {Mendez},
  \& {Lopez}}]{dinescu1997}
{Dinescu}, D.~I., {Girard}, T.~M., {van Altena}, W.~F., {Mendez}, R.~A., \&
  {Lopez}, C.~E. 1997, \aj, 114, 1014

\bibitem[{{Dinescu} {et~al.}(1999{\natexlab{b}}){Dinescu}, {van Altena},
  {Girard}, \& {L{\'o}pez}}]{dinescu1999a}
{Dinescu}, D.~I., {van Altena}, W.~F., {Girard}, T.~M., \& {L{\'o}pez}, C.~E.
  1999{\natexlab{b}}, \aj, 117, 277

\bibitem[{{Dotter} {et~al.}(2008){Dotter}, {Chaboyer}, {Jevremovi{\'c}},
  {Kostov}, {Baron}, \& {Ferguson}}]{dotter2008}
{Dotter}, A., {Chaboyer}, B., {Jevremovi{\'c}}, D., {et~al.} 2008, \apjs, 178,
  89

\bibitem[{{Ferraro} {et~al.}(1999){Ferraro}, {Messineo}, {Fusi Pecci}, {de
  Palo}, {Straniero}, {Chieffi}, \& {Limongi}}]{ferraro1999}
{Ferraro}, F.~R., {Messineo}, M., {Fusi Pecci}, F., {et~al.} 1999, \aj, 118,
  1738

\bibitem[{{Freire} {et~al.}(2003){Freire}, {Camilo}, {Kramer}, {Lorimer},
  {Lyne}, {Manchester}, \& {D'Amico}}]{freire2003}
{Freire}, P.~C., {Camilo}, F., {Kramer}, M., {et~al.} 2003, \mnras, 340, 1359

\bibitem[{{Gaia Collaboration} {et~al.}(2016){Gaia Collaboration}, {Brown},
  {Vallenari}, {Prusti}, {de Bruijne}, {Mignard}, {Drimmel}, {Babusiaux},
  {Bailer-Jones}, {Bastian}, \& et~al.}]{gaia2016}
{Gaia Collaboration}, {Brown}, A.~G.~A., {Vallenari}, A., {et~al.} 2016, \aap,
  595, A2

\bibitem[{{Geffert}(1998)}]{geffert1998}
{Geffert}, M. 1998, \aap, 340, 305

\bibitem[{{Gratton} {et~al.}(2003){Gratton}, {Bragaglia}, {Carretta},
  {Clementini}, {Desidera}, {Grundahl}, \& {Lucatello}}]{gratton2003}
{Gratton}, R.~G., {Bragaglia}, A., {Carretta}, E., {et~al.} 2003, \aap, 408,
  529

\bibitem[{{Hansen} {et~al.}(2007){Hansen}, {Anderson}, {Brewer}, {Dotter},
  {Fahlman}, {Hurley}, {Kalirai}, {King}, {Reitzel}, {Richer}, {Rich}, {Shara},
  \& {Stetson}}]{hansen2007}
{Hansen}, B.~M.~S., {Anderson}, J., {Brewer}, J., {et~al.} 2007, \apj, 671, 380

\bibitem[{{Harris}(1996)}]{harris1996}
{Harris}, W.~E. 1996, \aj, 112, 1487

\bibitem[{{Hendricks} {et~al.}(2012){Hendricks}, {Stetson}, {VandenBerg}, \&
  {Dall'Ora}}]{hendricks2012}
{Hendricks}, B., {Stetson}, P.~B., {VandenBerg}, D.~A., \& {Dall'Ora}, M. 2012,
  \aj, 144, 25

\bibitem[{{Heyl} {et~al.}(2012){Heyl}, {Richer}, {Anderson}, {Fahlman},
  {Dotter}, {Hurley}, {Kalirai}, {Rich}, {Shara}, {Stetson}, {Woodley}, \&
  {Zurek}}]{heyl2012}
{Heyl}, J.~S., {Richer}, H., {Anderson}, J., {et~al.} 2012, \apj, 761, 51

\bibitem[{{H{\o}g} {et~al.}(2000){H{\o}g}, {Fabricius}, {Makarov}, {Urban},
  {Corbin}, {Wycoff}, {Bastian}, {Schwekendiek}, \& {Wicenec}}]{hog2000}
{H{\o}g}, E., {Fabricius}, C., {Makarov}, V.~V., {et~al.} 2000, \aap, 355, L27

\bibitem[{{Kalirai} {et~al.}(2004){Kalirai}, {Richer}, {Hansen}, {Stetson},
  {Shara}, {Saviane}, {Rich}, {Limongi}, {Ibata}, {Gibson}, {Fahlman}, \&
  {Brewer}}]{kalirai2004}
{Kalirai}, J.~S., {Richer}, H.~B., {Hansen}, B.~M., {et~al.} 2004, \apj, 601,
  277

\bibitem[{{Kalirai} {et~al.}(2007){Kalirai}, {Anderson}, {Richer}, {King},
  {Brewer}, {Carraro}, {Davis}, {Fahlman}, {Hansen}, {Hurley}, {L{\'e}pine},
  {Reitzel}, {Rich}, {Shara}, \& {Stetson}}]{kalirai2007}
{Kalirai}, J.~S., {Anderson}, J., {Richer}, H.~B., {et~al.} 2007, \apjl, 657,
  L93

\bibitem[{{Kallivayalil} {et~al.}(2006){Kallivayalil}, {van der Marel}, \&
  {Alcock}}]{kallivayalil2006}
{Kallivayalil}, N., {van der Marel}, R.~P., \& {Alcock}, C. 2006, \apj, 652,
  1213

\bibitem[{{Kaluzny} {et~al.}(2013){Kaluzny}, {Thompson}, {Rozyczka}, {Dotter},
  {Krzeminski}, {Pych}, {Rucinski}, {Burley}, \& {Shectman}}]{kaluzny2013}
{Kaluzny}, J., {Thompson}, I.~B., {Rozyczka}, M., {et~al.} 2013, \aj, 145, 43

\bibitem[{{Krauss} \& {Chaboyer}(2003)}]{krauss2003}
{Krauss}, L.~M., \& {Chaboyer}, B. 2003, Science, 299, 65

\bibitem[{{Kroupa} \& {Bastian}(1997)}]{kroupa1997}
{Kroupa}, P., \& {Bastian}, U. 1997, \na, 2, 77

\bibitem[{{Kunder} {et~al.}(2013){Kunder}, {Stetson}, {Cassisi}, {Layden},
  {Bono}, {Catelan}, {Walker}, {Paredes Alvarez}, {Clem}, {Matsunaga},
  {Salaris}, {Lee}, \& {Chaboyer}}]{kunder2013}
{Kunder}, A., {Stetson}, P.~B., {Cassisi}, S., {et~al.} 2013, \aj, 146, 119

\bibitem[{{Kunder} {et~al.}(2014){Kunder}, {Bono}, {Piffl}, {Steinmetz},
  {Grebel}, {Anguiano}, {Freeman}, {Kordopatis}, {Zwitter}, {Scholz}, {Gibson},
  {Bland-Hawthorn}, {Seabroke}, {Boeche}, {Siebert}, {Wyse}, {Bienaym{\'e}},
  {Navarro}, {Siviero}, {Minchev}, {Parker}, {Reid}, {Gilmore}, {Munari}, \&
  {Helmi}}]{kunder2014}
{Kunder}, A., {Bono}, G., {Piffl}, T., {et~al.} 2014, \aap, 572, A30

\bibitem[{{Kunder} {et~al.}(2017){Kunder}, {Kordopatis}, {Steinmetz},
  {Zwitter}, {McMillan}, {Casagrande}, {Enke}, {Wojno}, {Valentini},
  {Chiappini}, {Matijevi{\v c}}, {Siviero}, {de Laverny}, {Recio-Blanco},
  {Bijaoui}, {Wyse}, {Binney}, {Grebel}, {Helmi}, {Jofre}, {Antoja}, {Gilmore},
  {Siebert}, {Famaey}, {Bienaym{\'e}}, {Gibson}, {Freeman}, {Navarro},
  {Munari}, {Seabroke}, {Anguiano}, {{\v Z}erjal}, {Minchev}, {Reid},
  {Bland-Hawthorn}, {Kos}, {Sharma}, {Watson}, {Parker}, {Scholz}, {Burton},
  {Cass}, {Hartley}, {Fiegert}, {Stupar}, {Ritter}, {Hawkins}, {Gerhard},
  {Chaplin}, {Davies}, {Elsworth}, {Lund}, {Miglio}, \& {Mosser}}]{kunder2016}
{Kunder}, A., {Kordopatis}, G., {Steinmetz}, M., {et~al.} 2017, \aj, 153, 75

\bibitem[{{K{\"u}pper} {et~al.}(2015){K{\"u}pper}, {Balbinot}, {Bonaca},
  {Johnston}, {Hogg}, {Kroupa}, \& {Santiago}}]{kuepper2015}
{K{\"u}pper}, A.~H.~W., {Balbinot}, E., {Bonaca}, A., {et~al.} 2015, \apj, 803,
  80

\bibitem[{{Lee} {et~al.}(2009){Lee}, {Kang}, {Lee}, \& {Lee}}]{lee2009}
{Lee}, J.-W., {Kang}, Y.-W., {Lee}, J., \& {Lee}, Y.-W. 2009, \nat, 462, 480

\bibitem[{{Lindegren} {et~al.}(2016){Lindegren}, {Lammers}, {Bastian},
  {Hern{\'a}ndez}, {Klioner}, {Hobbs}, {Bombrun}, {Michalik}, {Ramos-Lerate},
  {Butkevich}, {Comoretto}, {Joliet}, {Holl}, {Hutton}, {Parsons},
  {Steidelm{\"u}ller}, {Abbas}, {Altmann}, {Andrei}, {Anton}, {Bach},
  {Barache}, {Becciani}, {Berthier}, {Bianchi}, {Biermann}, {Bouquillon},
  {Bourda}, {Br{\"u}semeister}, {Bucciarelli}, {Busonero}, {Carlucci},
  {Casta{\~n}eda}, {Charlot}, {Clotet}, {Crosta}, {Davidson}, {de Felice},
  {Drimmel}, {Fabricius}, {Fienga}, {Figueras}, {Fraile}, {Gai}, {Garralda},
  {Geyer}, {Gonz{\'a}lez-Vidal}, {Guerra}, {Hambly}, {Hauser}, {Jordan},
  {Lattanzi}, {Lenhardt}, {Liao}, {L{\"o}ffler}, {McMillan}, {Mignard}, {Mora},
  {Morbidelli}, {Portell}, {Riva}, {Sarasso}, {Serraller}, {Siddiqui}, {Smart},
  {Spagna}, {Stampa}, {Steele}, {Taris}, {Torra}, {van Reeven}, {Vecchiato},
  {Zschocke}, {de Bruijne}, {Gracia}, {Raison}, {Lister}, {Marchant},
  {Messineo}, {Soffel}, {Osorio}, {de Torres}, \& {O'Mullane}}]{lindegren2016}
{Lindegren}, L., {Lammers}, U., {Bastian}, U., {et~al.} 2016, \aap, 595, A4

\bibitem[{{Mackey} \& {Gilmore}(2004)}]{mackey2004}
{Mackey}, A.~D., \& {Gilmore}, G.~F. 2004, \mnras, 355, 504

\bibitem[{{Marconi} {et~al.}(2003){Marconi}, {Caputo}, {Di Criscienzo}, \&
  {Castellani}}]{marconi2003}
{Marconi}, M., {Caputo}, F., {Di Criscienzo}, M., \& {Castellani}, M. 2003,
  \apj, 596, 299

\bibitem[{{Marino} {et~al.}(2011){Marino}, {Sneden}, {Kraft}, {Wallerstein},
  {Norris}, {da Costa}, {Milone}, {Ivans}, {Gonzalez}, {Fulbright}, {Hilker},
  {Piotto}, {Zoccali}, \& {Stetson}}]{marino2011}
{Marino}, A.~F., {Sneden}, C., {Kraft}, R.~P., {et~al.} 2011, \aap, 532, A8

\bibitem[{{Massari} {et~al.}(2017){Massari}, {Posti}, {Helmi}, {Fiorentino}, \&
  {Tolstoy}}]{massari2017}
{Massari}, D., {Posti}, L., {Helmi}, A., {Fiorentino}, G., \& {Tolstoy}, E.
  2017, \aap, 598, L9

\bibitem[{{McLaughlin} \& {van der Marel}(2005)}]{mclaughlin2005}
{McLaughlin}, D.~E., \& {van der Marel}, R.~P. 2005, \apjs, 161, 304

\bibitem[{{McMillan}(2011)}]{mcmillan2011}
{McMillan}, P.~J. 2011, \mnras, 414, 2446

\bibitem[{{Michalik} {et~al.}(2015){Michalik}, {Lindegren}, \&
  {Hobbs}}]{michalik2015}
{Michalik}, D., {Lindegren}, L., \& {Hobbs}, D. 2015, \aap, 574, A115

\bibitem[{{Milone} {et~al.}(2006){Milone}, {Villanova}, {Bedin}, {Piotto},
  {Carraro}, {Anderson}, {King}, \& {Zaggia}}]{milone2006}
{Milone}, A.~P., {Villanova}, S., {Bedin}, L.~R., {et~al.} 2006, \aap, 456, 517

\bibitem[{{Monaco} {et~al.}(2004){Monaco}, {Pancino}, {Ferraro}, \&
  {Bellazzini}}]{monaco2004}
{Monaco}, L., {Pancino}, E., {Ferraro}, F.~R., \& {Bellazzini}, M. 2004,
  \mnras, 349, 1278

\bibitem[{{Neeley} {et~al.}(2015){Neeley}, {Marengo}, {Bono}, {Braga},
  {Dall'Ora}, {Stetson}, {Buonanno}, {Ferraro}, {Freedman}, {Iannicola},
  {Madore}, {Matsunaga}, {Monson}, {Persson}, {Scowcroft}, \&
  {Seibert}}]{neeley2015}
{Neeley}, J.~R., {Marengo}, M., {Bono}, G., {et~al.} 2015, \apj, 808, 11

\bibitem[{{Ness} {et~al.}(2013){Ness}, {Freeman}, {Athanassoula},
  {Wylie-de-Boer}, {Bland-Hawthorn}, {Asplund}, {Lewis}, {Yong}, {Lane},
  {Kiss}, \& {Ibata}}]{ness2013}
{Ness}, M., {Freeman}, K., {Athanassoula}, E., {et~al.} 2013, \mnras, 432, 2092

\bibitem[{{Norris} \& {Freeman}(1983)}]{norris1983}
{Norris}, J., \& {Freeman}, K.~C. 1983, \apj, 266, 130

\bibitem[{{Odenkirchen} {et~al.}(1997){Odenkirchen}, {Brosche}, {Geffert}, \&
  {Tucholke}}]{odenkirchen1997}
{Odenkirchen}, M., {Brosche}, P., {Geffert}, M., \& {Tucholke}, H.-J. 1997,
  \na, 2, 477

\bibitem[{{Paez} {et~al.}(1990){Paez}, {Martinez Roger}, \&
  {Straniero}}]{paez1990}
{Paez}, E., {Martinez Roger}, C., \& {Straniero}, O. 1990, \aaps, 84, 481

\bibitem[{{Pearson} {et~al.}(2015){Pearson}, {K{\"u}pper}, {Johnston}, \&
  {Price-Whelan}}]{pearson2015}
{Pearson}, S., {K{\"u}pper}, A.~H.~W., {Johnston}, K.~V., \& {Price-Whelan},
  A.~M. 2015, \apj, 799, 28

\bibitem[{{Peterson} \& {Cudworth}(1994)}]{peterson1994}
{Peterson}, R.~C., \& {Cudworth}, K.~M. 1994, \apj, 420, 612

\bibitem[{{Peterson} {et~al.}(1995){Peterson}, {Rees}, \&
  {Cudworth}}]{peterson1995}
{Peterson}, R.~C., {Rees}, R.~F., \& {Cudworth}, K.~M. 1995, \apj, 443, 124

\bibitem[{{Reid} \& {Gizis}(1998)}]{reid1998}
{Reid}, I.~N., \& {Gizis}, J.~E. 1998, \aj, 116, 2929

\bibitem[{{Robin} {et~al.}(2003){Robin}, {Reyl{\'e}}, {Derri{\`e}re}, \&
  {Picaud}}]{robin2003}
{Robin}, A.~C., {Reyl{\'e}}, C., {Derri{\`e}re}, S., \& {Picaud}, S. 2003,
  \aap, 409, 523

\bibitem[{{Salaris} {et~al.}(2007){Salaris}, {Held}, {Ortolani}, {Gullieuszik},
  \& {Momany}}]{salaris2007}
{Salaris}, M., {Held}, E.~V., {Ortolani}, S., {Gullieuszik}, M., \& {Momany},
  Y. 2007, \aap, 476, 243

\bibitem[{{Sandage} \& {Cacciari}(1990)}]{sandage1990}
{Sandage}, A., \& {Cacciari}, C. 1990, \apj, 350, 645

\bibitem[{{Scholz} {et~al.}(1993){Scholz}, {Odenkirchen}, \&
  {Irwin}}]{scholz1993}
{Scholz}, R.~D., {Odenkirchen}, M., \& {Irwin}, M.~J. 1993, \mnras, 264, 579

\bibitem[{{Sch{\"o}nrich} {et~al.}(2010){Sch{\"o}nrich}, {Binney}, \&
  {Dehnen}}]{schoenrich2010}
{Sch{\"o}nrich}, R., {Binney}, J., \& {Dehnen}, W. 2010, \mnras, 403, 1829

\bibitem[{{Sirianni} {et~al.}(2005){Sirianni}, {Jee}, {Ben{\'{\i}}tez},
  {Blakeslee}, {Martel}, {Meurer}, {Clampin}, {De Marchi}, {Ford}, {Gilliland},
  {Hartig}, {Illingworth}, {Mack}, \& {McCann}}]{sirianni2005}
{Sirianni}, M., {Jee}, M.~J., {Ben{\'{\i}}tez}, N., {et~al.} 2005, \pasp, 117,
  1049

\bibitem[{{Thompson} {et~al.}(2010){Thompson}, {Kaluzny}, {Rucinski},
  {Krzeminski}, {Pych}, {Dotter}, \& {Burley}}]{thompson2010}
{Thompson}, I.~B., {Kaluzny}, J., {Rucinski}, S.~M., {et~al.} 2010, \aj, 139,
  329

\bibitem[{{van der Marel} \& {Sahlmann}(2016)}]{vandermarel2016}
{van der Marel}, R.~P., \& {Sahlmann}, J. 2016, \apjl, 832, L23

\bibitem[{{Verde} {et~al.}(2013){Verde}, {Jimenez}, \& {Feeney}}]{verde2013}
{Verde}, L., {Jimenez}, R., \& {Feeney}, S. 2013, Physics of the Dark Universe,
  2, 65

\bibitem[{{Watkins} {et~al.}(2015{\natexlab{a}}){Watkins}, {van der Marel},
  {Bellini}, \& {Anderson}}]{watkins2015b}
{Watkins}, L.~L., {van der Marel}, R.~P., {Bellini}, A., \& {Anderson}, J.
  2015{\natexlab{a}}, \apj, 812, 149

\bibitem[{{Watkins} {et~al.}(2015{\natexlab{b}}){Watkins}, {van der Marel},
  {Bellini}, \& {Anderson}}]{watkins2015a}
---. 2015{\natexlab{b}}, \apj, 803, 29

\bibitem[{{Webb} {et~al.}(2014){Webb}, {Leigh}, {Sills}, {Harris}, \&
  {Hurley}}]{webb2014}
{Webb}, J.~J., {Leigh}, N., {Sills}, A., {Harris}, W.~E., \& {Hurley}, J.~R.
  2014, \mnras, 442, 1569

\bibitem[{{Woodley} {et~al.}(2012){Woodley}, {Goldsbury}, {Kalirai}, {Richer},
  {Tremblay}, {Anderson}, {Bergeron}, {Dotter}, {Esteves}, {Fahlman}, {Hansen},
  {Heyl}, {Hurley}, {Rich}, {Shara}, \& {Stetson}}]{woodley2012}
{Woodley}, K.~A., {Goldsbury}, R., {Kalirai}, J.~S., {et~al.} 2012, \aj, 143,
  50

\bibitem[{{Wu} {et~al.}(2002){Wu}, {Wang}, \& {Chen}}]{wu2002}
{Wu}, Z.-Y., {Wang}, J.-J., \& {Chen}, L. 2002, \cjaa, 2, 216

\bibitem[{{Zhu} {et~al.}(2016){Zhu}, {Romanowsky}, {van de Ven}, {Long},
  {Watkins}, {Pota}, {Napolitano}, {Forbes}, {Brodie}, \& {Foster}}]{zhu2016}
{Zhu}, L., {Romanowsky}, A.~J., {van de Ven}, G., {et~al.} 2016, \mnras, 462,
  4001

\bibitem[{{Zinn}(1985)}]{zinn1985}
{Zinn}, R. 1985, \apj, 293, 424

\bibitem[{{Zocchi} {et~al.}(2016){Zocchi}, {Gieles}, {H{\'e}nault-Brunet}, \&
  {Varri}}]{zocchi2016}
{Zocchi}, A., {Gieles}, M., {H{\'e}nault-Brunet}, V., \& {Varri}, A.~L. 2016,
  \mnras, 462, 696

\end{thebibliography}


\appendix

\section{Additional Figures}
\label{sect:extrafigs}

\begin{figure*}[!b]
    \centering
    \includegraphics[width=0.24\linewidth]{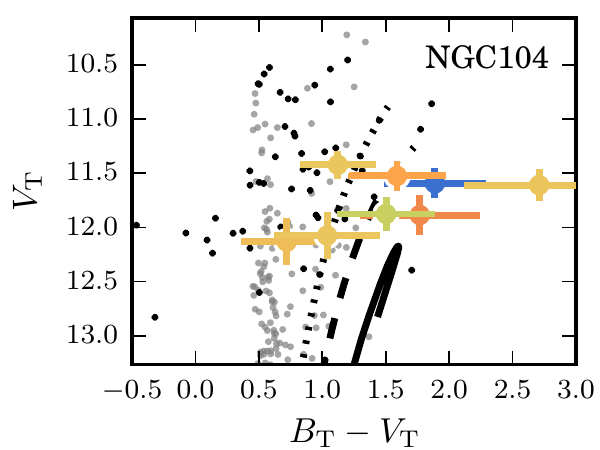}
    \includegraphics[width=0.24\linewidth]{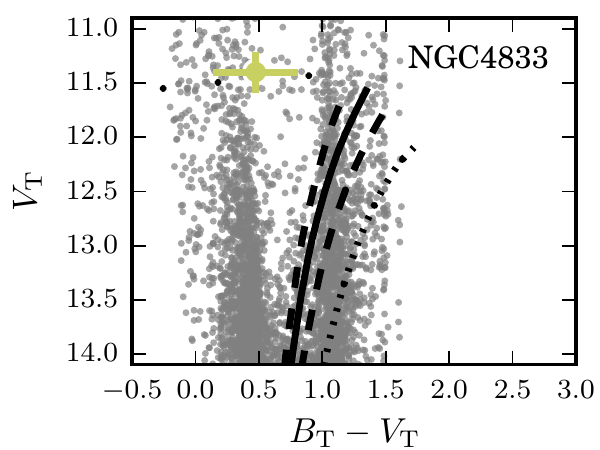}
    \includegraphics[width=0.24\linewidth]{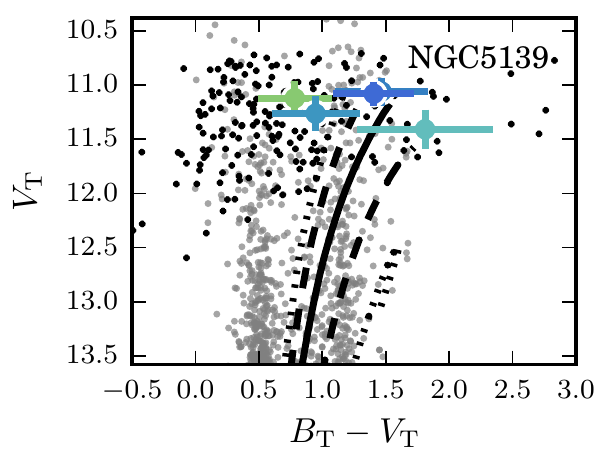}
    \includegraphics[width=0.24\linewidth]{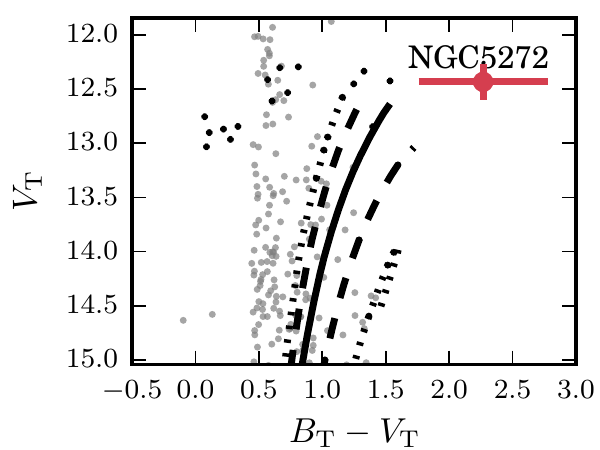}
    \includegraphics[width=0.24\linewidth]{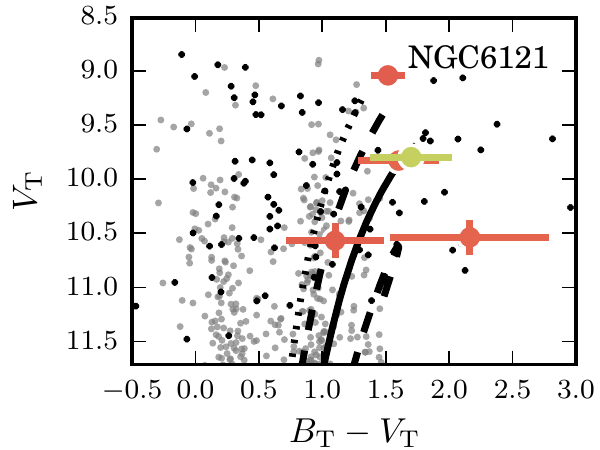}
    \includegraphics[width=0.24\linewidth]{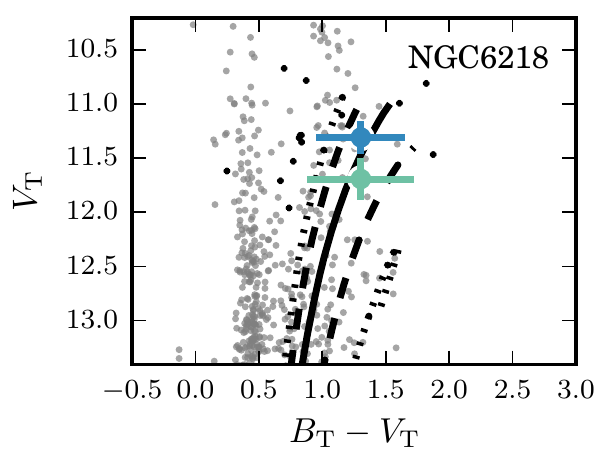}
    \includegraphics[width=0.24\linewidth]{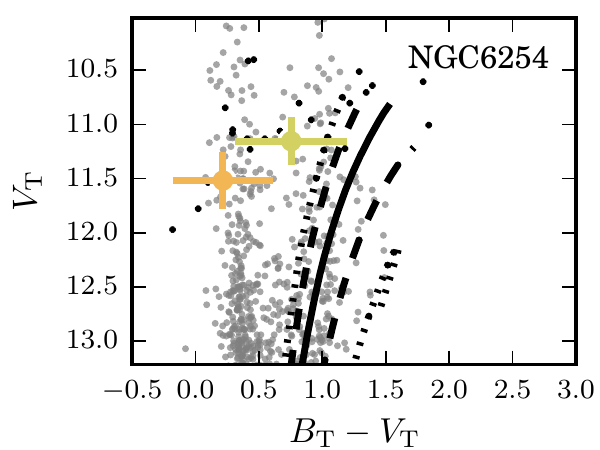}
    \includegraphics[width=0.24\linewidth]{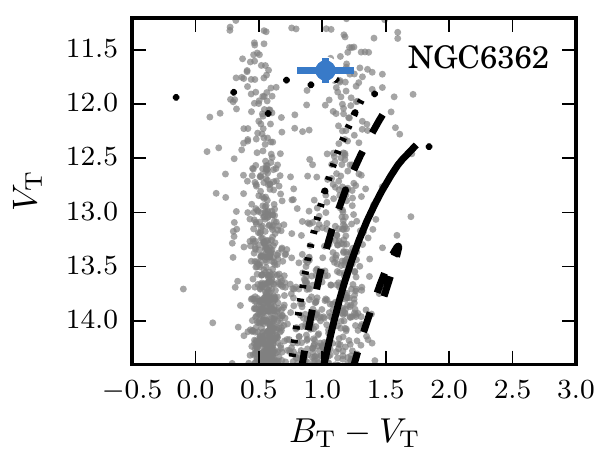}
    \includegraphics[width=0.24\linewidth]{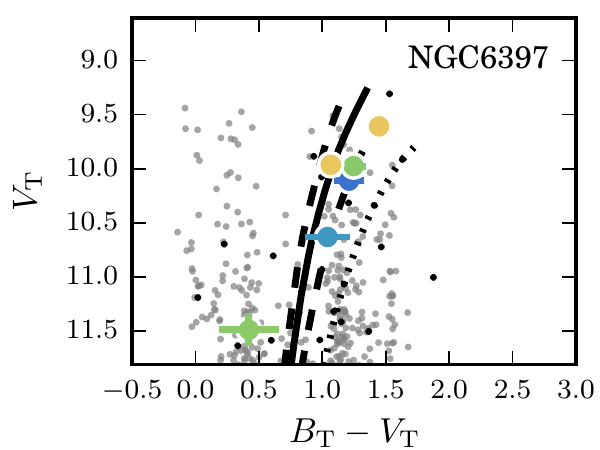}
    \includegraphics[width=0.24\linewidth]{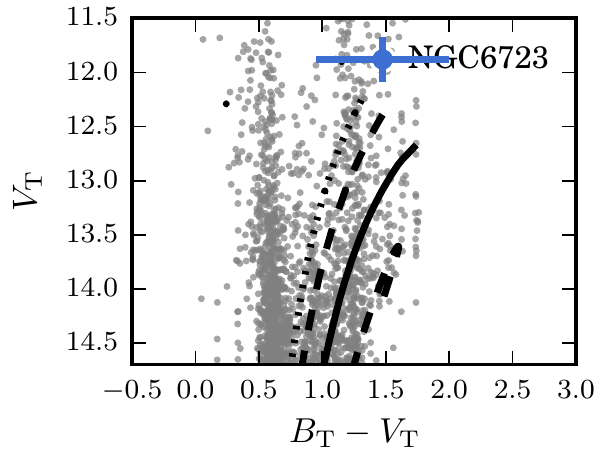}
    \includegraphics[width=0.24\linewidth]{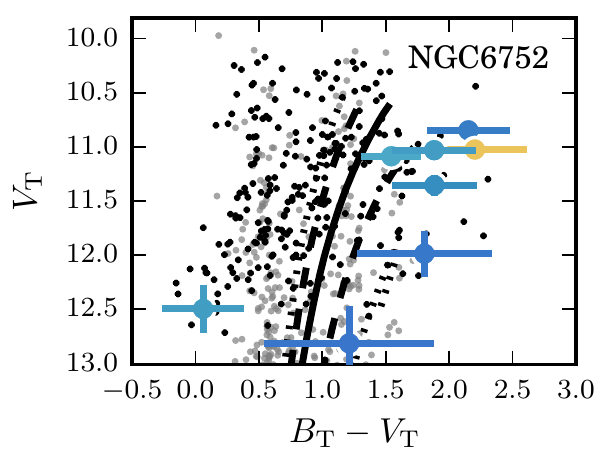}
    \includegraphics[width=0.24\linewidth]{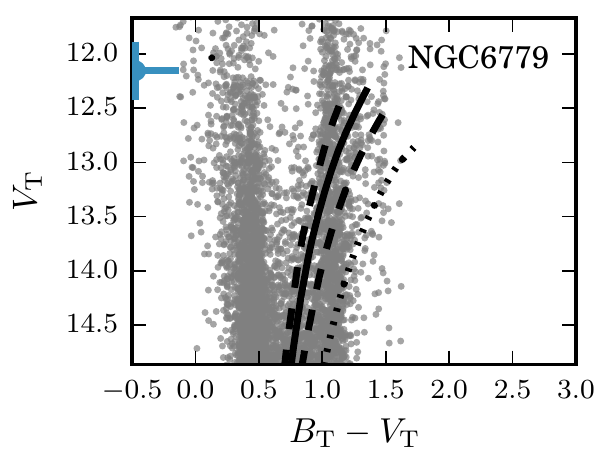}
    \includegraphics[width=0.24\linewidth]{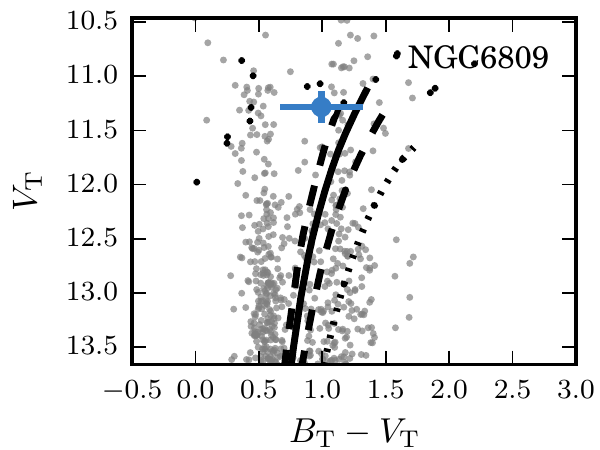}
    \includegraphics[width=0.24\linewidth]{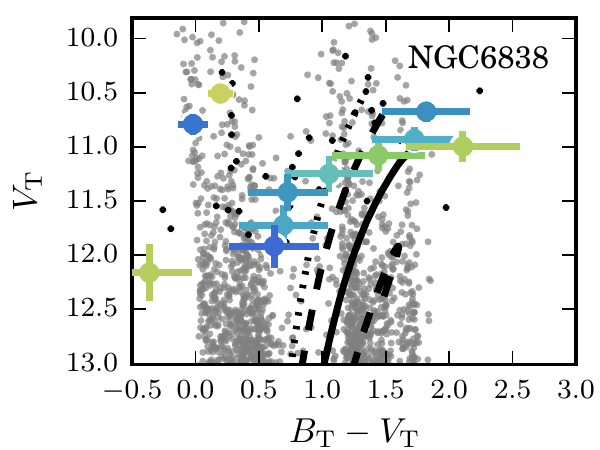}
    \caption{Same as \autoref{fig:cmds} for NGC\,104, NGC\,4833, NGC\,5139, NGC\,5272, NGC\,6121, NGC\,6218, NGC\,6254, NGC\,6362, NGC\,6397, NGC\,6723, NGC\,6752, NGC\,6779, NGC\,6809, and NGC\,6838.}
    \label{fig:cmds_app}
\end{figure*}

In \autoref{ssect:photometry}, we determined whether stars identified as candidate cluster members are photometrically consistent with their nearest cluster. We showed the results of this analysis for NGC\,6656 in \autoref{fig:cmds} as an example. \autoref{fig:cmds_app} shows the results for the other 14 clusters for which we performed this analysis.

\begin{figure*}
    \centering
    \includegraphics[width=0.49\linewidth]{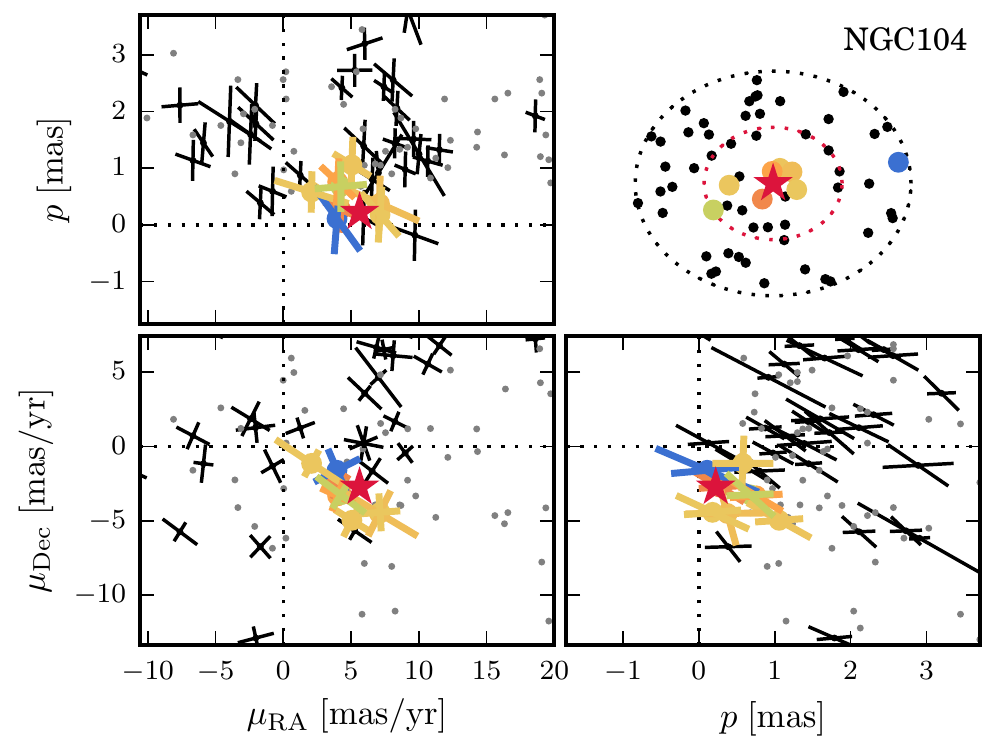}
    \includegraphics[width=0.49\linewidth]{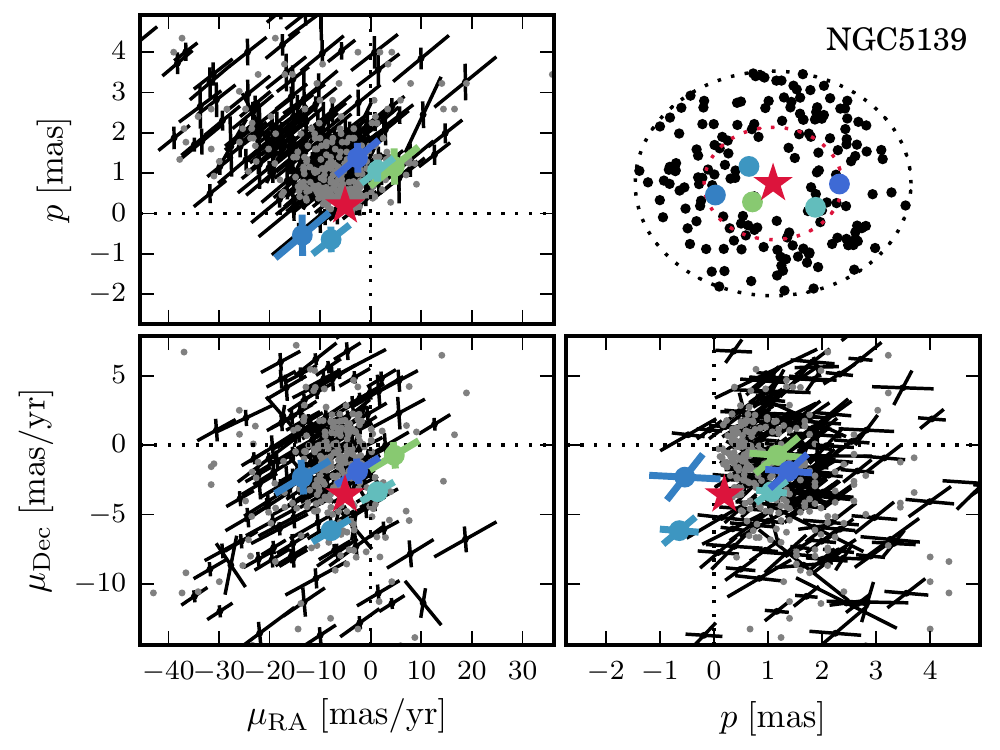}
    \includegraphics[width=0.49\linewidth]{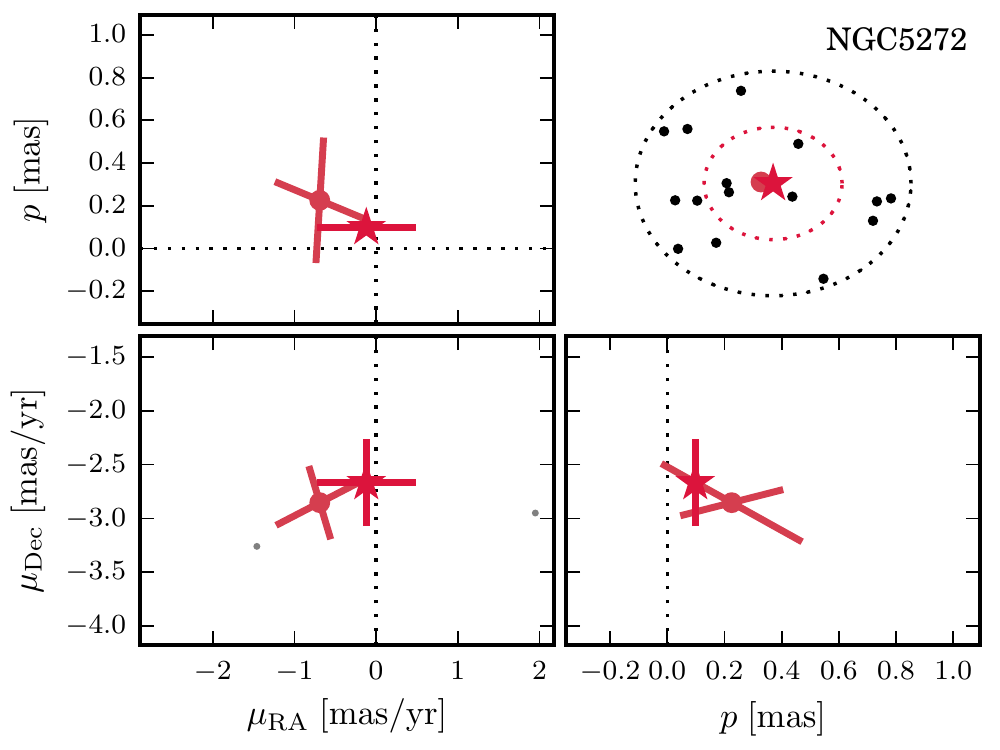}
    \includegraphics[width=0.49\linewidth]{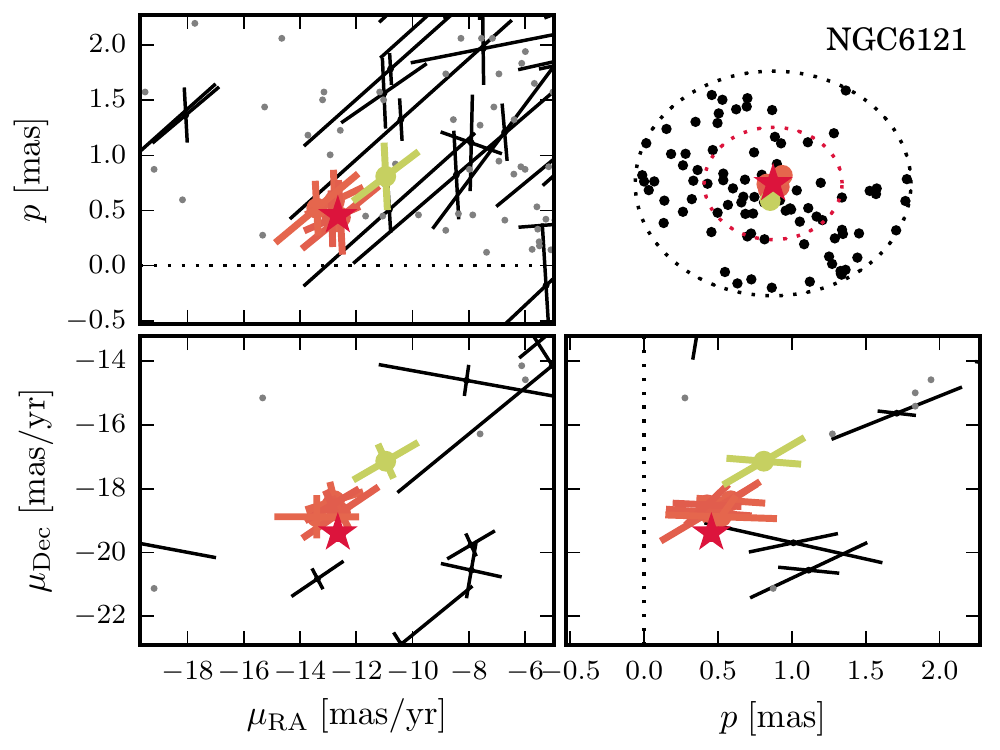}
    \includegraphics[width=0.49\linewidth]{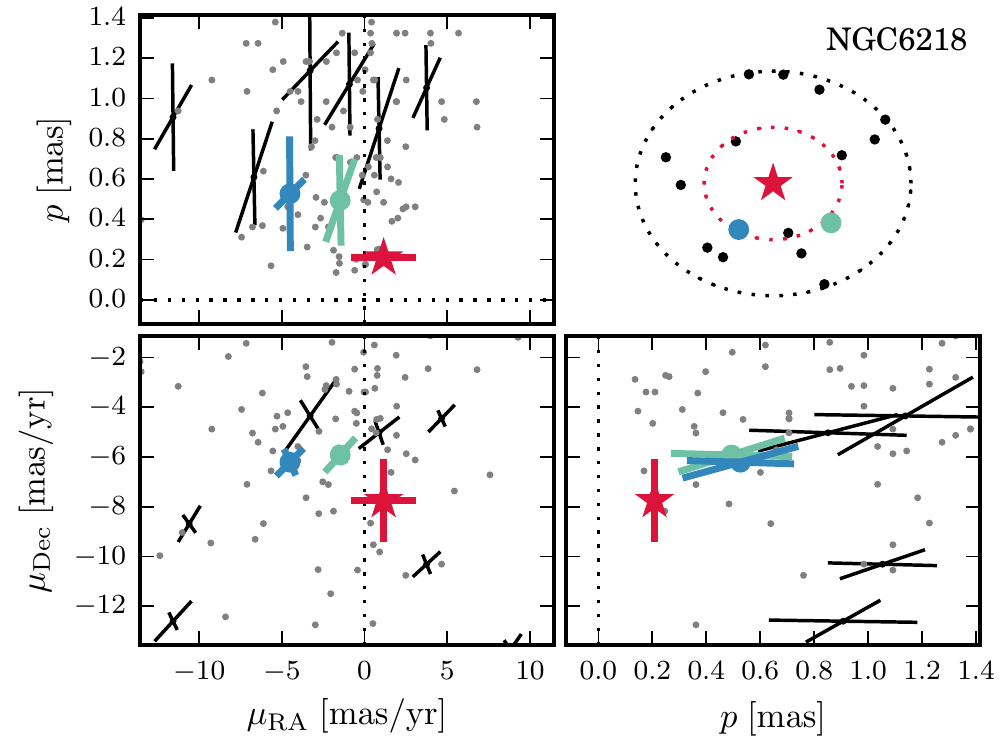}
    \includegraphics[width=0.49\linewidth]{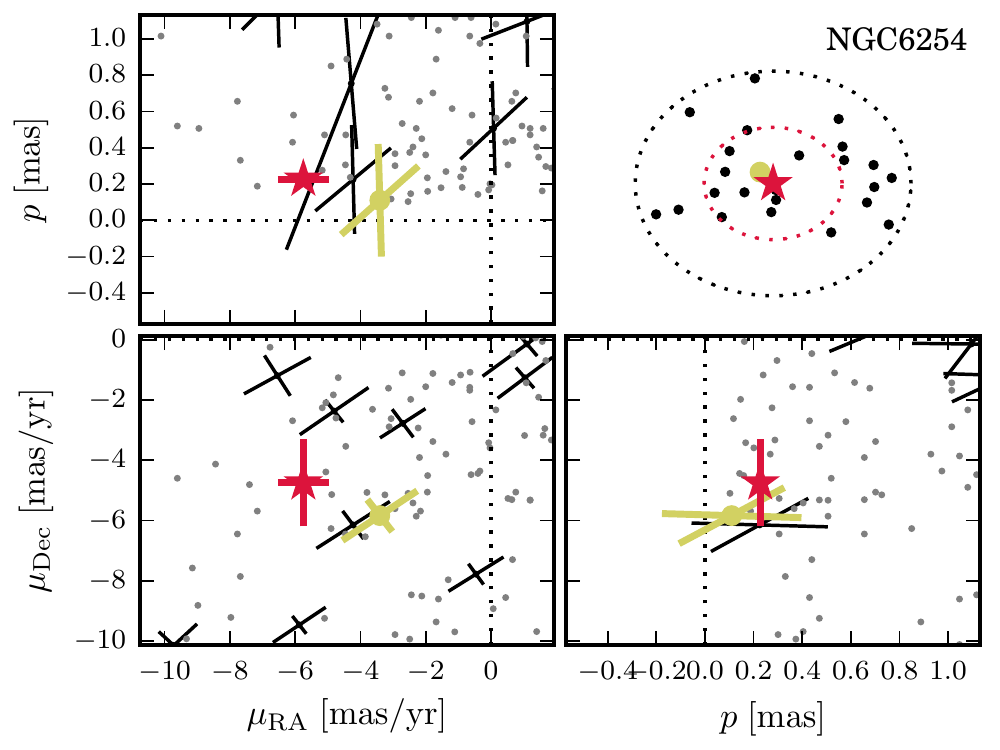}
    \caption{Same as \autoref{fig:probs} for NGC\,104, NGC\,5139, NGC\,5272, NGC\,6121, NGC\,6218, and NGC\,6254.}
    \label{fig:probs1_app}
\end{figure*}

\begin{figure*}
    \centering
    \includegraphics[width=0.49\linewidth]{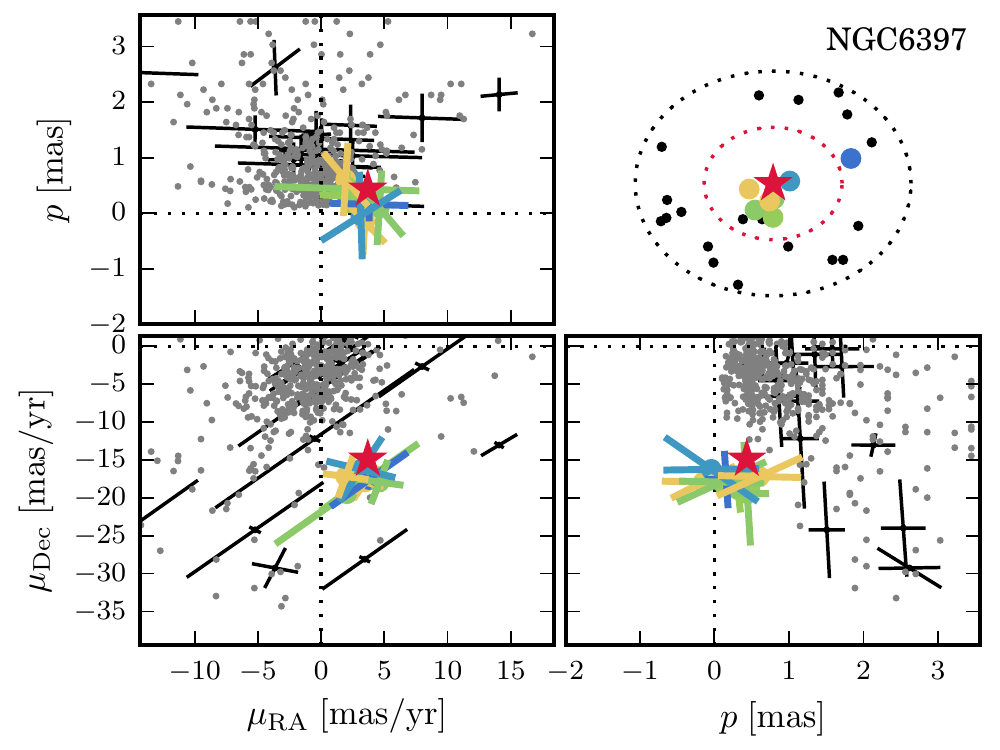}
    \includegraphics[width=0.49\linewidth]{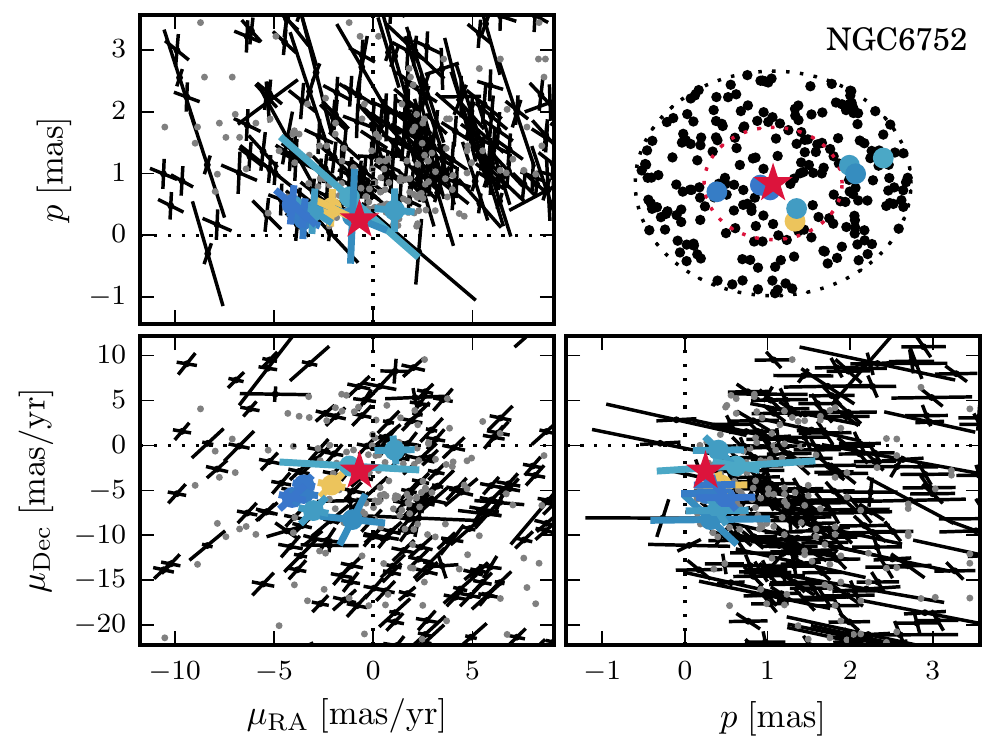}
    \includegraphics[width=0.49\linewidth]{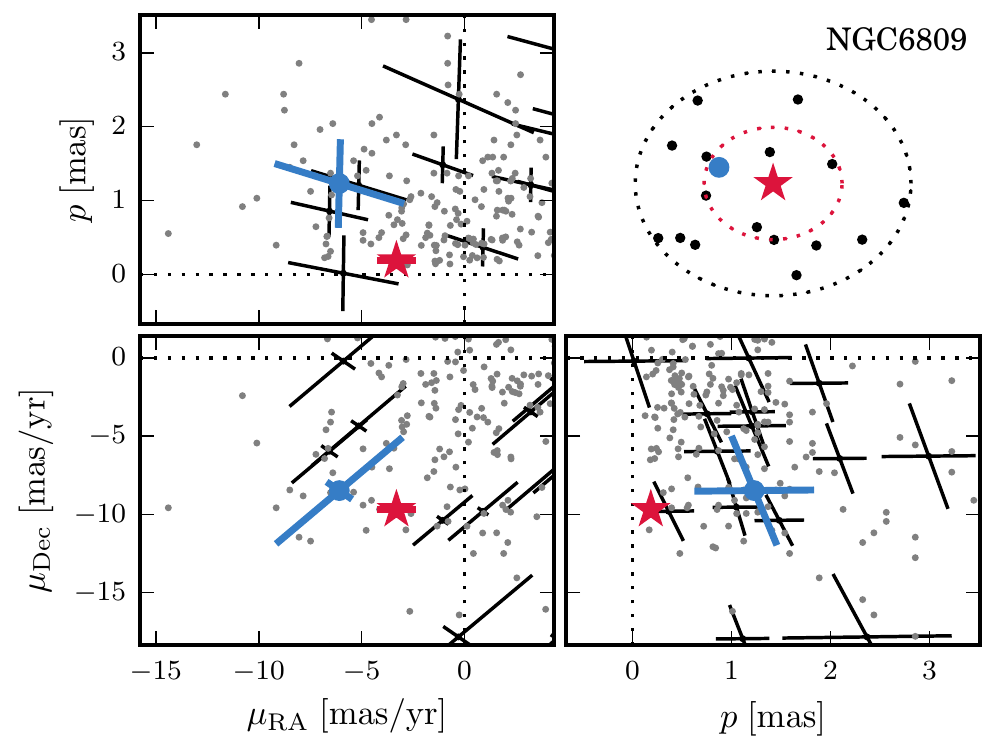}
    \includegraphics[width=0.49\linewidth]{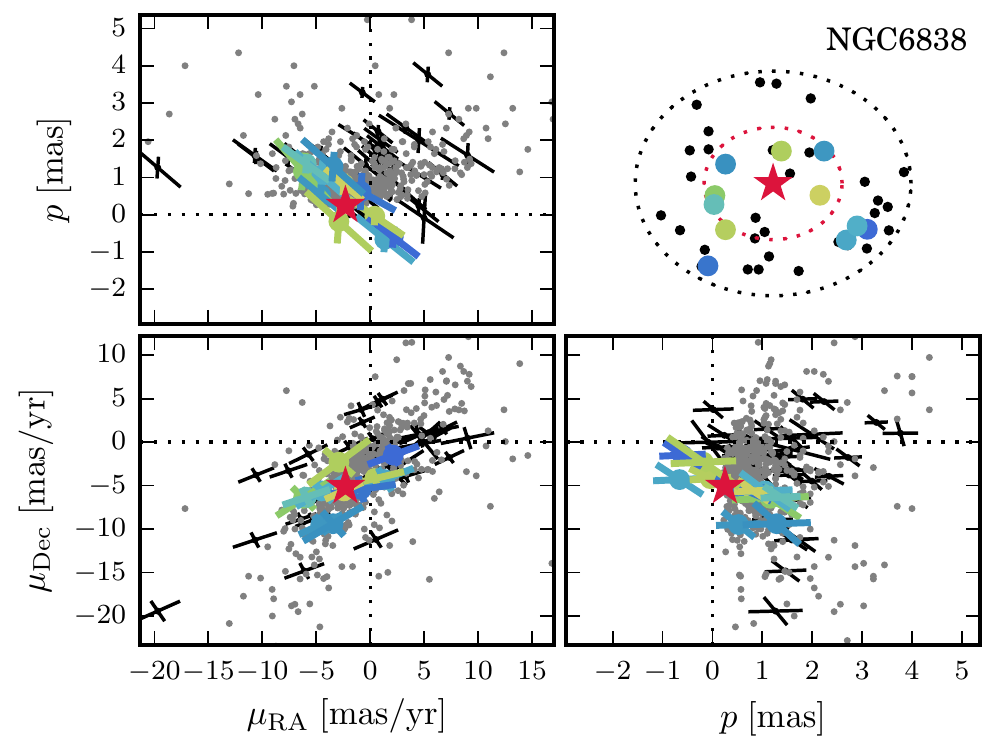}
    \caption{Same as \autoref{fig:probs} for NGC\,6397, NGC\,6752, NGC\,6809, and NGC\,6838.}
    \label{fig:probs2_app}
\end{figure*}

In \autoref{ssect:besancon}, we determined whether stars identified as candidate cluster members are consistent with predictions for the foreground population from the Besan\c{c}on simulations. We showed the results of this analysis for NGC\,6656 in \autoref{fig:probs} as an example. \autoref{fig:probs1_app} and \autoref{fig:probs2_app} show the results for the other 10 clusters for which we performed this analysis.

\end{document}